\documentclass[aip,amsmath,amssymb,reprint,]{revtex4-1}
\usepackage[colorlinks=true,citecolor=blue,linkcolor=blue,urlcolor=blue]{hyperref}
\usepackage{graphicx}
\usepackage{dcolumn}
\usepackage{bm}
\usepackage[utf8]{inputenc}
\usepackage[T1]{fontenc}
\usepackage{mathptmx}
\usepackage{etoolbox}
\makeatletter
\def\@email#1#2{
 \endgroup
 \patchcmd{\titleblock@produce}
  {\frontmatter@RRAPformat}
  {\frontmatter@RRAPformat{\produce@RRAP{*#1\href{mailto:#2}{#2}}}\frontmatter@RRAPformat}
  {}{}
}
\makeatother
\begin{document}

\preprint{AIP/123-QED}

\title[Phase formation in hole- and electron-doped rare-earth nickelate single crystals]{Phase formation in hole- and electron-doped rare-earth nickelate single crystals}
\author{P. Puphal*}
\email{puphal@fkf.mpg.de}
\author{V. Sundaramurthy}
\author{V. Zimmermann}
\author{K. Küster}
\author{U. Starke}
\author{M. Isobe}
\author{B. Keimer}
\author{M. Hepting}
\affiliation{Max Planck Institute for Solid State Research, Heisenbergstra{\ss}e 1, D-70569 Stuttgart, Germany}
\date{\today}

\begin{abstract}
The recent discovery of superconductivity in hole-doped infinite-layer nickelates has triggered a great interest in the synthesis of novel nickelate phases, which have primarily been examined in thin film samples. Here, we report the high-pressure optical floating zone (OFZ) growth of various perovskite and perovskite-derived rare-earth nickelate single-crystals, and investigate the effects of hole-, electron-, and self-doping. For hole-doping with Ca and Sr, we observe phase separations during the growth process when a substitution level of 8\% is exceeded. A similar trend emerges for electron-doping with Ce and Zr. Employing lower doping levels allows us to grow sizeable crystals in the perovskite phase, which exhibit significantly different electronic and magnetic properties than the undoped parent compounds, such as a decreased resistivity and a suppressed magnetic response. Our insights into the doping-dependent phase formation and the resulting properties of the synthesized crystals reveal limitations and opportunities for the exploration and manipulation of electronic states in rare-earth nickelates.

\end{abstract}
\maketitle

\begin{quotation}

\end{quotation}

\section{\label{sec:level1}Introduction}

The Ruddlesden-Popper nickelates $R_{n+1}$Ni$_n$O$_{3n+1}$ ($R$ = rare-earth ion) with the perovskite ($n = \infty$) or perovskite-derived crystal structures ($n \neq \infty$) are prototypical quantum materials \cite{Greenblatt1997}, showing a plethora of electronic, orbital, and magnetic phases already without charge carrier doping \cite{Sugai1990,Kobayashi1996,Mazin2007,liu2021d,Catalano2018}. The perovskite-derived phases with $n = 1 - 3$ have been actively studied since the 1980s \cite{Petrov1988}. To synthesize the perovskite phase $R$NiO$_3$, high pressure is required to stabilize the Ni$^{3+}$ oxidation state and the distorted crystal structure where the octahedral tilt angles are determined by the radius of the $R$ ion. The least distorted perovskite nickelate LaNiO$_{3}$ was first synthesized in 1957 \cite{Wold1957}, whereas nickelates with $R$ other than La were realized in 1991 \cite{Lacorre1991}. The $R$NiO$_3$ compounds have been of long-standing interest due to their sharp metal-to-insulator transition \cite{Torrance1992,Medarde1992,Catalan2008} and a magnetic ground state with an unconventional spin spiral \cite{Scagnoli2006,Hepting2018}. Additionally, the metal-to-insulator transition is accompanied by an orthorhombic to monoclinic structural phase transition \cite{Catalan2008}. The transition temperatures of these phases decrease as the size of the $R$ ion increases. One exception is the compound with the largest anion, LaNiO$_{3}$, which exhibits a rhombohedral structure and remains metallic and paramagnetic down to the lowest temperatures, provided that the oxygen content is stoichiometric \cite{Wang2018,Zheng2020}.

Upon hole- or electron-doping of the Ruddlesden-Popper nickelate, new electronic phases can emerge \cite{Iglesias2021,Botana2017,Kotiuga2019,Worm2022,Chen2023,Patel2022}, along with potential functional properties \cite{Li2022,Song2023}. In $R$NiO$_3$ powders, hole- and electron doping up to 10\% has been achieved through the substitution of the $R$ ion by divalent Sr/Ca and tetravalent Th/Ce ions, respectively \cite{Cheong1994,Alonso1995,GarciaMunoz1995,Ramadoss2016}. Typically, the powder synthesis is carried out under high external oxygen gas pressures \cite{Lacorre1991}. Recent studies have reported substitutions in powders as high as 40\% \cite{Li2020}. However, further investigations are needed to determine if such high substitution concentrations are homogeneously incorporated into the microstructure of the $R$NiO$_3$ powder grains.

Since 2019, the field has been reinvigorated \cite{Mitchell2021} by the discovery of superconducting behavior in hole-doped rare-earth nickelates with the infinite-layer crystal structure \cite{Li2019,Zeng2020,Lee2020,Osada2021,Gao2021,zeng2021}. This structure can be achieved through the topotactic oxygen deintercalation of the perovskite phase \cite{Crespin1983,Hayward1999,Hayward2003}. These infinite-layer nickelates with Ni$^{1+}$ are nominally isoelectronic and isostructural to cuprate high-temperature superconductors with Cu$^{2+}$ ions \cite{Hepting2021}. Additionally, superconductivity has been observed in the undoped nickelate Nd$_{6}$Ni$_{5}$O$_{12}$ \cite{Pan2021}, which, however, can be considered a self-doped cuprate analogue \cite{Botana2022}. Yet, superconductivity in the material family of nickelates has only been observed in thin film samples \cite{Wang20201,Li2020}, even though the latest studies have indicated that it could be an intrinsic property of the bulk phase of doped infinite-layer nickelates \cite{Goodge2023}. Furthermore, theoretical studies propose that superconductivity may also arise in electron-doped nickelates, such as La$_{2.4}$Zr$_{0.6}$Ni$_{2}$O$_{6}$ where La$^{3+}$ is partially substituted by tetravalent Zr$^{4+}$ ions \cite{Worm2022}.

Currently, a direct synthesis of the infinite-layer phase of nickelates is unfeasible, due to the highly metastable Ni$^{1+}$ state in square-planar NiO$_2$ units. Similarly, the direct synthesis of other Ruddlesden-Popper derived nickelates with NiO$_2$ planes, such as Nd$_{6}$Ni$_{5}$O$_{12}$, has yet to be achieved. However, well-established routes exist for the topochemical removal of the apical oxygen from the parent nickelate phases, such as using H$_2$/Ar gas, or CaH$_2$ powder as a reducing agent \cite{Crespin1983,Zhang2017,Pan2021}. In particular, the CaH$_2$-assisted reduction has been successfully demonstrated not only on perovskite thin films \cite{Li2019,Zeng2020,Lee2020,Osada2021,Gao2021,zeng2021} and polycrystalline powders \cite{Wang20201,Puphal2022,Ortiz2022}, but also on single-crystals \cite{Puphal2021,Wu2023} with volumes up to 1 mm$^3$ \cite{Puphal2023}. 

\begin{table*}[tb]
\caption{Summary of the materials without doping grown with the HKZ. The columns indicate the desired nominal composition of the material, $n$ labels the formed Ruddlesden-Popper (RP) phase, wt\% is the weight percentage of the RP phase, and $a, b, c$ are the lattice constants determined from Rietveld refinements of PXRD data. Furthermore, the space group (structure) and the oxygen partial pressure of the growth are indicated. If only one wt\% value is given, the remaining phase is NiO.}
\begin{tabular}{llllllll}
Nominal comp. & $n$ in RP & wt\% &  a ($\textrm{Å}$) & b ($\textrm{Å}$) & c ($\textrm{Å}$)  & structure & p(O${_2}$) (bar)  \tabularnewline \hline 
LaNiO$_{3}$ & $\infty$ & 97.9(7)  & 5.45408(7) & - & 13.1255(2)  & $R\bar3c$ & 85\tabularnewline 
PrNiO$_{3}$ & $\infty$ & 97.3(6)  & 5.4095(1) & 5.3703(1) & 7.6147(2)  & $Pbnm$ & 300\tabularnewline 
Pr$_{6}$Ni$_{5}$O$_{16}$ & 3 & 72.5(6)  & 5.3717(2)  & 5.4565(1)  & 27.534(1)  & $P2_{1}/a$ & 300\tabularnewline 
~~~~~~~~~~~``& 2 & 27.5(3) &  5.4406(3) & 5.3763(3) & 7.6166(4)  & $Pbnm$ & 300\tabularnewline 
La$_{3}$Eu$_{3}$Ni$_{5}$O$_{16}$ & 3 & 74.9(7) & 5.3657(2) & 5.4599(2) & 27.4703(9)  & $P2_{1}/a$ & 300\tabularnewline 
~~~~~~~~~~~``& 1 & 11.1(3) & 3.808(2)  & -  & 12.515(1)  & $I4/mmm$ & 300\tabularnewline 
\end{tabular}
\label{pure}
\end{table*}


\begin{table*}[tb]
\caption{Summary of the materials with doping grown with the HKZ. The columns
indicate the desired nominal composition of the material, $n$ labels the formed Ruddlesden-Popper (RP) phase (where DP stands for double perovskite), wt\% is the weight percentage of the RP phase, $x$ is the substitution content determined by EDS, and $a, b, c$ are the lattice constants determined from Rietveld refinements of PXRD data. Furthermore, the space group (structure) and the oxygen partial pressure of the growth are indicated. If only one wt\% value is given, the remaining phase is NiO. The labels (in), (mid), and (out) indicate whether the powder for the PXRD analysis was extracted from the inner, middle, or outer part of the boule, as depicted in Fig. \ref{LSNO}.}
\begin{tabular}{lllllllll}
Nominal comp.  & $n$ in RP & wt\%  & $x$ from EDS  & a ($\textrm{Å}$)  & b ($\textrm{Å}$)  & c ($\textrm{\textrm{Å}}$)    & structure  & p(O${_{2}}$) (bar) \tabularnewline
\hline 
Pr$_{0.95}$Ce$_{0.05}$NiO$_{3}$  & $\infty$ & 41.3(3)  & 0.02(1)  & 5.4097(1)  & 5.3748(2)  & 7.6201(3)    & $Pbnm$  & 300\tabularnewline
\hline 
La$_{0.95}$Ce$_{0.05}$NiO$_{3}$  & $\infty$  & 77.5(4) & 0.046(3)  & 5.45350(8)  & - & 13.1235(2)   & $R\bar{3}c$  & 85\tabularnewline
La$_{0.91}$Ce$_{0.09}$NiO$_{3}$  & $\infty$ & 85.81(5)  & 0.057(7)  & 5.4530(1)  & - & 13.1201(3)   & $R\bar{3}c$  & 85\tabularnewline
La$_{0.8}$Ce$_{0.2}$NiO$_{3}$  & $\infty$ & 69.2(4)  & 0.07(1)  & 5.4532(1)  & - & 13.1193(3)   & $R\bar{3}c$  & 200\tabularnewline
\hline 
La$_{2.4}$Zr$_{0.6}$Ni$_{2}$O$_{7}$  & $\infty$ & 46.14(4) & 0.12(1)  & 5.5019(2)  &  & 13.2531(5)    & $R\bar{3}c$  & 14\tabularnewline
~~~~~~~~~~~~~~`` & DP & 52.6(5)  & - & 5.6231(2)  & 5.7053(3)  & 7.9891(4)    & $P2_{1}/n$  & 14\tabularnewline
La$_{2.4}$Zr$_{0.6}$Ni$_{2}$O$_{7}$  & DP & 60.5(7) & - & 5.6404(2)  & 5.7442(3)  & 8.0209(4)    & $P2_{1}/n$  & 4.5\tabularnewline
~~~~~~~~~~~~~~`` & 1 & 19.3(2) & - & 3.8637(1)  & - & 12.6857(7)    & $I4/mmm$  & 4.5 \tabularnewline
\hline 
La$_{0.98}$Zr$_{0.02}$NiO$_{3}$  & $\infty$  & 99.5(2) & 0.026(6)  & 5.48144(2)  & - & 13,18948(7)   & $R\bar{3}c$  & 85\tabularnewline
La$_{0.95}$Zr$_{0.05}$NiO$_{3}$  & $\infty$ & 99.5(2) & 0.071(4)  & 5.48144(2)  & - & 13,18948(7)    & $R\bar{3}c$  & 85\tabularnewline
La$_{0.8}$Zr$_{0.2}$NiO$_{3}$  & $\infty$ & 57.8(5) & 0.12(1)  & 5.5027(1)  & - & 13.2596(4)    & $R\bar{3}c$  & 30\tabularnewline
~~~~~~~~~~~~~~`` & DP & 39.39(4) & - & 5.6223(2)  & 5.6987(3)  & 7.9781(4)    & $P2_{1}/n$  & 30\tabularnewline
\hline 
La$_{0.92}$Sr$_{0.08}$NiO$_{3}$  & $\infty$  & 65.2(6)  & 0.12(5)  & 5.4471(2)  & - & 13.1691(1)   & $R\bar{3}c$  & 200\tabularnewline
La$_{0.84}$Sr$_{0.16}$NiO$_{3}$ (in)  & $\infty$ & 27.5(2) & 0.13(1)  & 5.44551(7)  & - & 13.1395(3)    & $R\bar{3}c$  & 300\tabularnewline
~~~~~~~~~~~~~~`` & 1 & 67.1(4)   & - & 3.82578(4)  & - & 12.7002(1)   & $I4/mmm$  & 300\tabularnewline
La$_{0.84}$Sr$_{0.16}$NiO$_{3}$ (mid)  & $\infty$ & 54.9(3) & - & 5.44372(6)  & - & 13.1341(2)    & $R\bar{3}c$  & 300\tabularnewline
~~~~~~~~~~~~~~`` & 1 & 31.7(2) & - & 3.81085(6)  & - & 12.7141(9)    & $I4/mmm$  & 300 \tabularnewline
La$_{0.84}$Sr$_{0.16}$NiO$_{3}$ (out)  & $\infty$ & 72.1(4)  & - & 5.44539(4)  & - & 13.1382(1)    & $R\bar{3}c$  & 300\tabularnewline
~~~~~~~~~~~~~~`` & 1 & 29.4(2) & - & 3.81860(5)  & - & 12.7172(3)    & $I4/mmm$  & 300 \tabularnewline
\hline 
\end{tabular}\label{refinement} 
\end{table*}

A technical breakthrough enabling the growth of large $R_{n+1}$Ni$_n$O$_{3n+1}$ single-crystals was the advent of the high-oxygen pressure high-temperature optical floating zone technique, yielding bulk single-crystals  of LaNiO$_{3}$ \cite{Guo2018,Wang2018,Zhang2017LNO,Zheng2020,Dey2019,Puphal2023}, PrNiO$_{3}$ \cite{Zheng2019}, La$_{3}$Ni$_{2}$O$_{7}$ \cite{Liu2022}, and (La,Pr)$_{4}$Ni$_{3}$O$_{10}$ \cite{Zhang2020c,Huangfu2020}. However, a problem with the oxygen transport towards the center of the boule was observed in this synthesis method \cite{Zheng2020}. As a result, oxygen deficient phases may form, but can be alleviated by post annealing under high gas pressure in autoclaves.

Overall, the technical advances in the perovskite single-crystal growth as well as the demonstrated topotactic reduction of LaNiO$_3$ single-crystals to infinite-layer LaNiO$_2$ \cite{Puphal2023}, call for the synthesis of doped perovskite single-crystals that can possibly be reduced to the infinite-layer phase. In this work we present the optical floating zone growth of single-crystals of doped perovskite and Ruddlesden-Popper phase nickelates under 300 bar of oxygen partial pressure, and discuss the opportunities and limitations of this method. We begin with an overview on the synthesis of the undoped compounds LaNiO$_{3}$ and PrNiO$_{3}$, as well as an attempt of the growth of the $n = 5$ Ruddlesden-Popper nickelate Pr$_{6}$Ni$_{5}$O$_{16}$. As a second type of samples, we investigate Ce- and Zr-doped (La,Pr)NiO$_{3}$, which are nominally electron-doped. We also report on Sr-doped LaNiO$_{3}$, which is nominally hole-doped.

\section{\label{sec:level2}Methods}

Precursor powders were prepared by mixing the corresponding stoichiometries of La$_{2}$O$_{3}$ (99.99\% Alfa Aesar), Pr$_{6}$O$_{11}$ (Alfa Aesar, 99.99\%) and NiO (99.998\% Alfa Aesar), as well as CeO$_{2}$ (99.99\% Alfa Aesar), CaCO$_{3}$ (99.999\% Alfa Aesar), SrCO$_{3}$ (99.998\% Roth), ZrO$_{2}$ (99.978\% Alfa Aesar), and Eu$_{2}$O$_{3}$ (99.995\% Roth). Subsequently, the powders were ball-milled for 20 minutes and the mixtures were transferred in alumina crucibles to box furnaces, followed by heating to 1100$^{\circ}$C. 
Cylindrically shaped feed and seed rods were prepared by ball-milling the sintered materials, which were filled into rubber forms with 6~mm diameter. The rubber was evacuated and pressed in a stainless steel form filled with water using a Riken type S1-120 70 kN press. All rods were heat treated at 900$^{\circ}$C to avoid cracking or breaking due to the oxidization process, when the density of the rod becomes too high.

The single-crystal growth was carried out in a high pressure, high-temperature, optical floating zone furnace (model HKZ, SciDre GmbH, Dresden, Germany), that allows for gas pressures in the growth chamber up to 300 bar. The growth chamber has a length of 72~mm and 20~mm wall thickness. A xenon arc lamp operating at 5 kW was used as a heating source with the rare vertical mirror alignment of the HKZ. The rods were then aligned in the HKZ on steel holders followed by the installation of the high pressure chamber. Subsequently the chamber was filled up to 14/ 30/ 85/ 200/ 300 bar oxygen pressure and held at a flow rate of 0.1 l/min. The crucial part in the nickelate growth is the initial melt connection which is complicated, as we typically start with a mixture of the Ruddlesden-Popper phase (La,Pr)$_{2}$NiO$_{4}$ and NiO, which both have lower melting points than the desired perovskite phase. However, a preannealing at high pressures leads to cracking and even breaking of the rods. Thus, we either increased the power until a homogeneous zone melt was achieved, or premelted the rods at the given pressure to preform the final phase. In the former case, achieving a homogenous zone melt is necessary before connecting the two rods, but the starting process of the growth was often very challenging. Both procedures have been employed for the materials synthesized in this work, and have proven to be equally successful. Notably, a premelting of the rods does not increase the oxygen content in the obtained boule, as the given oxygen pressure and oxygen transport in the melt stabilizes this, and the subsequent growth is easier. 

All growths were conducted under oxidizing atmosphere, with the partial oxygen pressure tailored to yield the best outcome for each composition (see Tab. \ref{pure}, \ref{refinement}). Common to all compounds investigated here is the increase in melting temperature when the phase containing Ni$^{3+}$ is formed. This effect is amplified by the endothermal oxidation process which we counteracted by a continuous increase of the lamp power, particularly during the early stage of the growth. This means that newly introduced material will be largely overheated and becomes very fluid. Thus, without premelting, the diameter of the grown boule is dictated by the flow of the low-viscosity melt that requires constant feeding. For all growths, we found that a growth rate of 2 mm/h with an additional feed rate of 2 mm/h results in stable conditions.

Powder x-ray diffraction (PXRD) was performed at room temperature using a Rigaku Miniflex diffractometer with Bragg Brentano geometry, Cu K$_\alpha$ radiation and a Ni filter. Rietveld refinements were conducted with the FullProf software suite \cite{Rodriguez1993}. 
The x-ray Laue diffraction images were collected with a Photonic Science CCD detector using a standard W broad x-ray source operated at 35 kV and 40 mA. For indexing of the Laue patterns, the software ORIENTEXPRESS was used.


Electron microscopy images with both secondary electrons (SE) and backscattering electrons (BSE) were taken with a Zeiss Merlin electron microscope operated at 12 kV, 600 mA at a sample distance of 5 mm. Energy-dispersive x-ray spectra  (EDS) were recorded with a NORAN System 7 (NSS212E) detector in a Tescan Vega (TS-5130MM) SEM.

Magnetic susceptibility measurements were performed using a vibrating sample magnetometer (MPMS VSM SQUID, Quantum Design) and electrical transport measurements with a Physical Property Measurement System (PPMS, Quantum Design).

X-ray photoelectron spectroscopy (XPS) data were collected using a commercial Kratos AXIS Ultra spectrometer and a monochromatized Al $K_\alpha$ source (photon energy, 1486.6 eV). The base pressure during XPS was in the low $10^{-10}$ mbar range. The spectra were collected using an analyzer pass energy of 20 eV. XPS spectra were analyzed using the CASAXPS software \cite{Fairley2021}. All samples were cleaved in a glove box, mounted on carbon tape or In foil, and transported under inert atmosphere to the XPS chamber.

\section{\label{sec:level3}Results}
\subsection{Undoped nickelates}
\begin{figure}[tb]
\includegraphics[width=1\columnwidth]{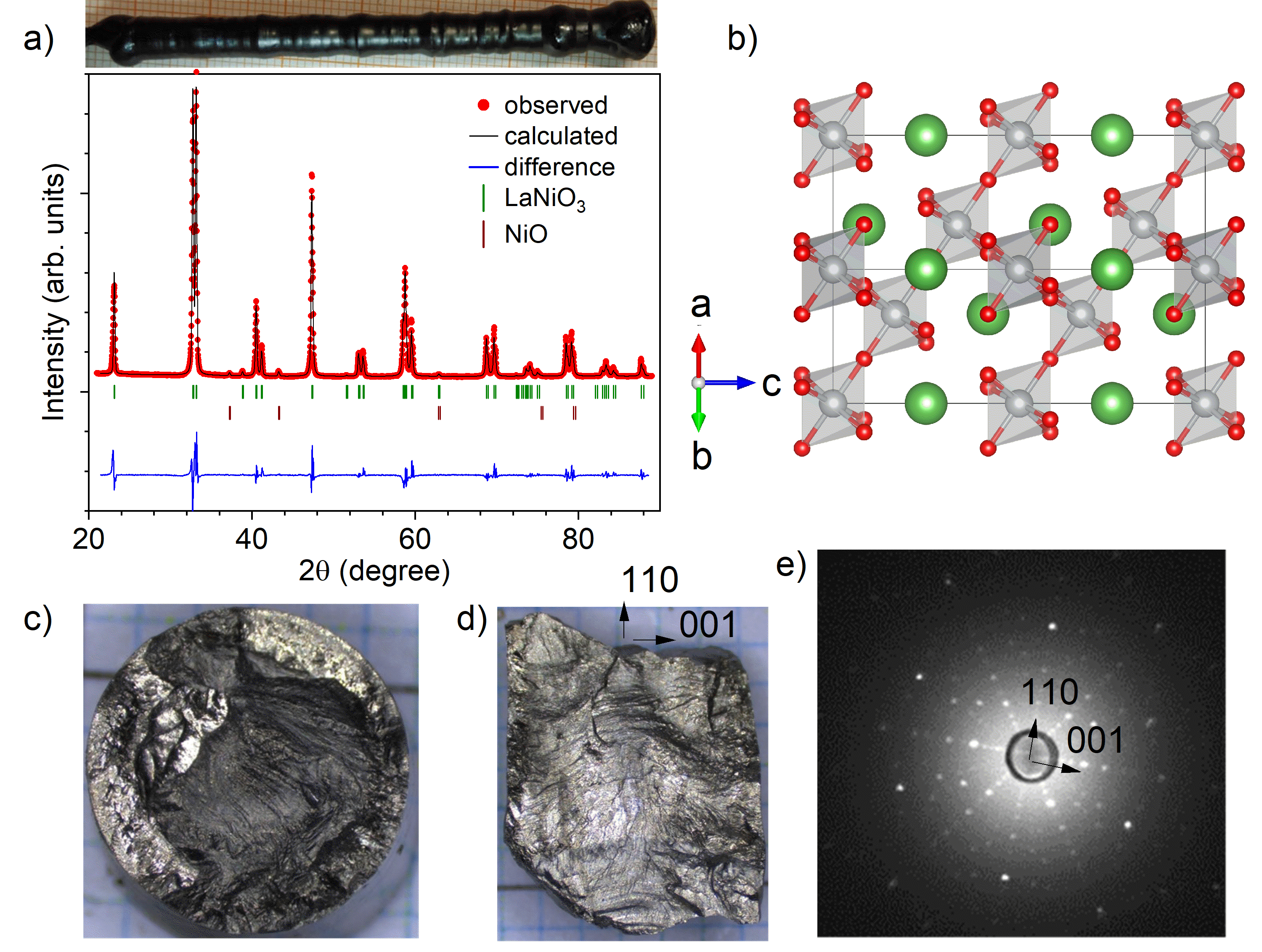}
\caption{(a) Image of the grown LaNiO$_{3}$ boule and the PXRD pattern. The solid black line corresponds to the calculated intensity from the Rietveld refinement. The calculated Bragg peak positions of the LaNiO$_{3}$ majority (97.9(7) wt\%) and NiO minority phase (2.1(2) wt\%) are indicated as green and red vertical bars, respectively. The blue line corresponds to the difference between the experimental and calculated intensity. (b) Schematic of the rhombohedral unit cell (hexagonal axes) of perovskite LaNiO$_{3}$, viewed along the same direction as the cleave shown in panel (d). (c) Picture of the cross section of the boule obtained after cleaving a single crystal. (d) Picture of a cleave along growth direction of a single crystalline piece with indexed orientations determined from the Laue diffraction pattern in panel (e). 
}
\label{LNO}
\end{figure}

\begin{figure*}[tb]
\includegraphics[width=1.8\columnwidth]{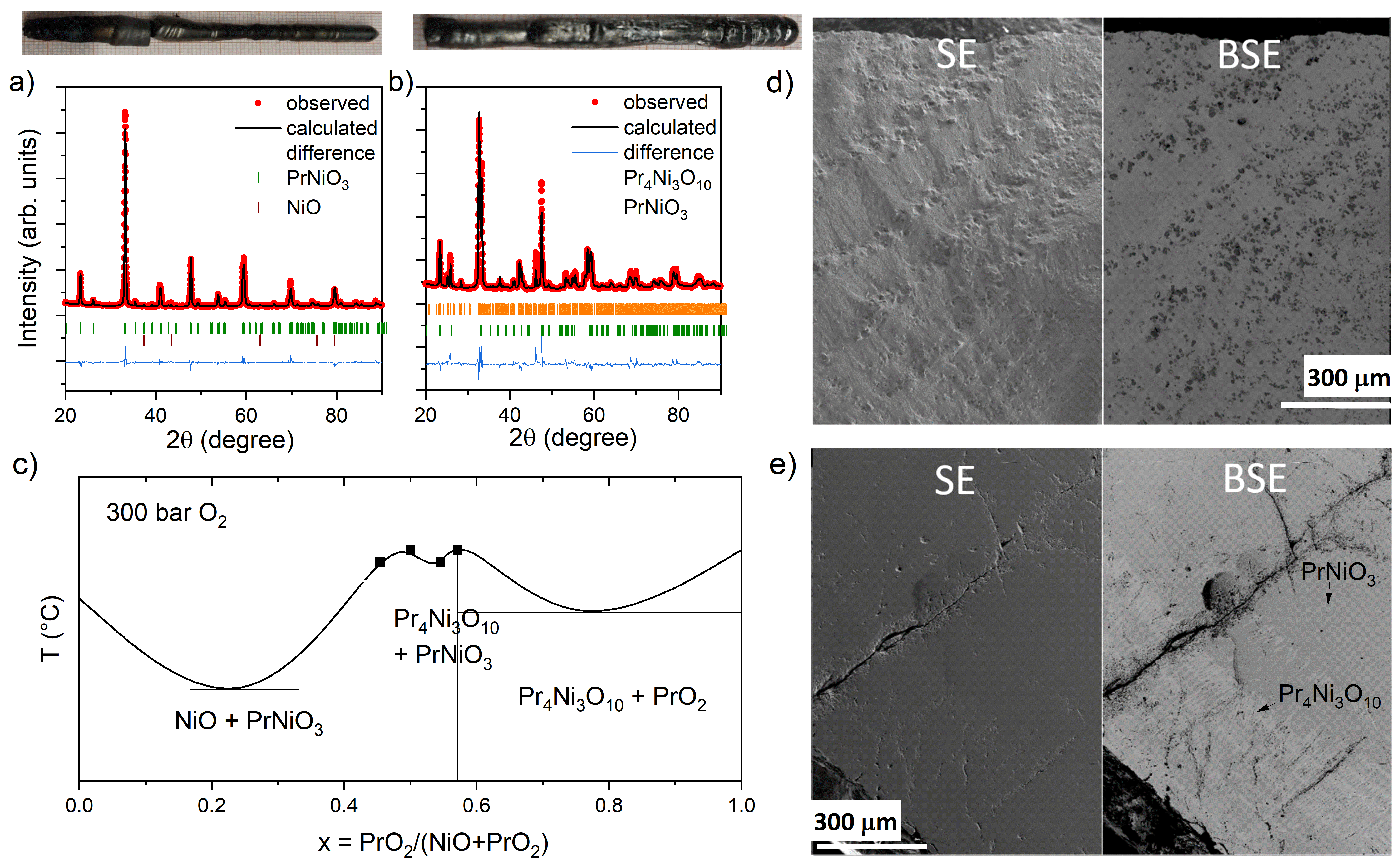}
\caption{(a) Image of the grown PrNiO$_3$ boule and the PXRD pattern. The calculated Bragg peak positions of the PrNiO$_3$ majority (97.3(7) wt\%) and NiO minority phase (2.7(2) wt\%) are indicated as green and purple vertical bars, respectively. (b) Image of the boule of the attempted Pr$_{6}$Ni$_{5}$O$_{16}$ growth and the PXRD pattern, which indicates only the Pr$_{4}$Ni$_{3}$O$_{10}$ (72.5(6) wt\%) and the PrNiO$_3$ phase (27.5(3) wt\%). (c) Sketch of the growth phase diagram for the phase formations of the OFZ melt under 300 bar oxygen partial pressure for NiO and PrO$_2$. (d,e) Simultaneously acquired SE (left) and BSE (right) images of broken surfaces from the boules grown from a precursor corresponding to nominally stoichiometric PrNiO$_3$ (d) and a Pr-excess precursor nominally corresponding to Pr$_{1.05}$NiO$_{3}$ (e). 
The inclusions with a dark contrast in (d) correspond to NiO, while regions with a light contrast in (e) correspond  to intergrown Pr$_{4}$Ni$_{3}$O$_{10}$.
}
\label{PNO}
\end{figure*}

Figure \ref{LNO}a shows the as-grown LaNiO$_{3}$ boule together with the characterization by PXRD and a Laue diffraction image (Fig.~\ref{LNO}e). The OFZ synthesis with a HKZ-type furnace of LaNiO$_{3}$ as well as PrNiO$_{3}$, La$_{3}$Ni$_{2}$O$_{7}$, and (La,Pr)$_{4}$Ni$_{3}$O$_{10}$ has been reported previously \cite{Guo2018,Wang2018,Zhang2017LNO,Zheng2020,Dey2019,Puphal2023,Zheng2019,Liu2022,Zhang2020c,Huangfu2020}, although only a few details were given about the growth, phase diagram and the formation of secondary phases. 
As can be seen in Fig. \ref{LNO}a, the PXRD reveals the presence of a small amount of NiO in addition to the primary LaNiO$_{3}$ phase in a pulverized piece broken off from the boule. The emergence of secondary phases such as NiO will be discussed in detail for the other nickelate compounds in this study. In accord with previous studies on LaNiO$_{3}$, we refine the PXRD data in the rhombohedral space group $R\bar3c$, which also describes the structure of the doped La-based perovskites in our study (see Tab. \ref{pure}, \ref{refinement}). As apparent from Laue diffraction (Fig. \ref{LNO}e), the growth direction of LaNiO$_{3}$ corresponds to the rhombohedral (110) direction. Under the application of a relatively strong force, the LaNiO$_{3}$ boule can be cleaved. The exposed surfaces are rough with small facets (Fig. \ref{LNO}c,d) that are mostly aligned in parallel to the (110) plane. The boule can also be cleaved along the orthogonal direction, which is displayed in Fig. \ref{LNO}d where the $(1\bar10)$ direction points out of the picture while the $c$-axis lies in the horizontal plane (see also the corresponding sketch of the crystal structure in Fig. \ref{LNO}b).

Figures \ref{PNO}a,b diplay the boules and the PXRD patterns of our PrNiO$_{3}$ growth and the attempted growth of Pr$_{6}$Ni$_{5}$O$_{16}$. To achieve a boule of PrNiO$_{3}$ with a length of about 8 cm (see Fig. \ref{PNO}a), a continuous pressure of 300 bar was held for several days. The results of our PXRD analysis of PrNiO$_{3}$ and our other growths are summarized in Tab.~ \ref{pure}, displaying the lattice constants and phase compositions in wt\%.
In agreement with previous studies, we refine the structure of PrNiO$_{3}$ in the orthorhombic space group $Pbnm$. In Fig. \ref{PNO}c we present the first phase diagram of NiO - PrO$_{2}$, which at the given pressure of 300 bar oxygen only contains four phases: NiO, PrNiO$_{3}$, Pr$_{4}$Ni$_{3}$O$_{10}$ and  PrO$_{2}$.

Similarly to LaNiO$_{3}$ (Fig. \ref{LNO}b), we detect a minor amount of NiO in the stoichiometrically grown PrNiO$_{3}$ (Fig. \ref{PNO}b). Note that an admixture of NiO to the perovskite phase can become relevant after a topotactic reduction where ferromagnetic Ni forms which can dominate the signal in magnetic susceptibility measurements \cite{Ortiz2022}. In cases of very small NiO admixture, its presence can be below the detection limit of standard laboratory PXRD. A method that allows to detect even subtle amounts of NiO inclusions is BSE imaging. In particular, unlike SE images that reflect the sample topography, BSE images reveal the spatial distribution of the elements. Figure \ref{PNO}d shows the simultaneously acquired SE and BSE images of a broken surface from the PrNiO$_{3}$ boule. In the latter image, NiO rich regions are clearly observed as the dark contrast.

In principle, the emergence of NiO rich regions should be alleviated in a growth from a precursor with Pr excess. However, we find that when varying the stoichiometry to a subtle Pr rich content, Pr$_{4}$Ni$_{3}$O$_{10}$ forms as an intergrown phase, especially in regions in proximity to the surface of the boule (see light contrast in the right panel of Fig. \ref{PNO}e). This occurs because impurity phases are usually pushed to the surface in an OFZ growth, as nucleation occurs in the core of the boule. Due to our external heating source the heat is only indirectly transported to the center resulting in a small gradient. Nonetheless this observation is in contrast to the pressure stability, as the center also sees the least pressure and Pr$_{4}$Ni$_{3}$O$_{10}$ is more stable at lower pressures.

For PrNiO$_{3}$, the growth direction corresponds to the orthorhombic (100) direction. Similarly to LaNiO$_{3}$, under the application of a relatively strong force, the PrNiO$_{3}$ boule can be cleaved. The exposed surfaces are rough with small facets that are mostly aligned in parallel to the (100) plane. The boule can also be cleaved along the orthogonal direction, where the $(010)$ direction points out of the plane and the (001) direction is the horizontal one. Notably, for both La- and Pr-compounds these cleaving patterns emerge even for the doped cases that will be discussed below.

Next, we turn to the synthesis of the $n=5$ Ruddlesden-Popper phase Pr$_{6}$Ni$_{5}$O$_{16}$ under 300 bar oxygen partial pressure. If successful, the compound could provide a perspective for topotactic reductions to the Pr$_{6}$Ni$_{5}$O$_{12}$ phase, in analogy to superconducting Nd$_{6}$Ni$_{5}$O$_{12}$ films \cite{Pan2021}, and seems ideally suited for the high-pressure OFZ growth, as the pressure stability range should fall into the region accessible with the HKZ. However, instead of Pr$_{6}$Ni$_{5}$O$_{16}$, our PXRD characterization reveals a phase mixture of the perovskite and the $n=3$ phase Pr$_{4}$Ni$_{3}$O$_{10}$ (Fig. \ref{PNO}b), which are the phases that encompass the $n=5$ compound in the growth phase diagram. Similarly, we find in an attempted synthesis (not shown here) that also the $n=5$ variant La$_{3}$Eu$_{3}$Ni$_{5}$O$_{16}$ cannot be stabilized under 300 bar oxygen (see Tab. \ref{pure}), in spite of the different ionic radii of La and Eu. This is distinct from other layered materials such as cuprates, where the mixing of ionic radii of the smaller rare earth ion with large Ba facilitates the stabilization of complex layered structures. Here we find that instead of the $n=5$, the $n=1$ and 3 Ruddlesden-Popper phases form. Hence, we conclude that at a given pressure of 300 bar, the synthesis of La- and Pr-based Ruddlesden-Popper phases higher than $n=3$ is likely unfeasible, as the existence of these phases is probably restricted to a very narrow range in pressure and composition space.

\subsection{Electron-doping of PrNiO$_{3}$}

\begin{figure}[tb]
\includegraphics[width=1\columnwidth]{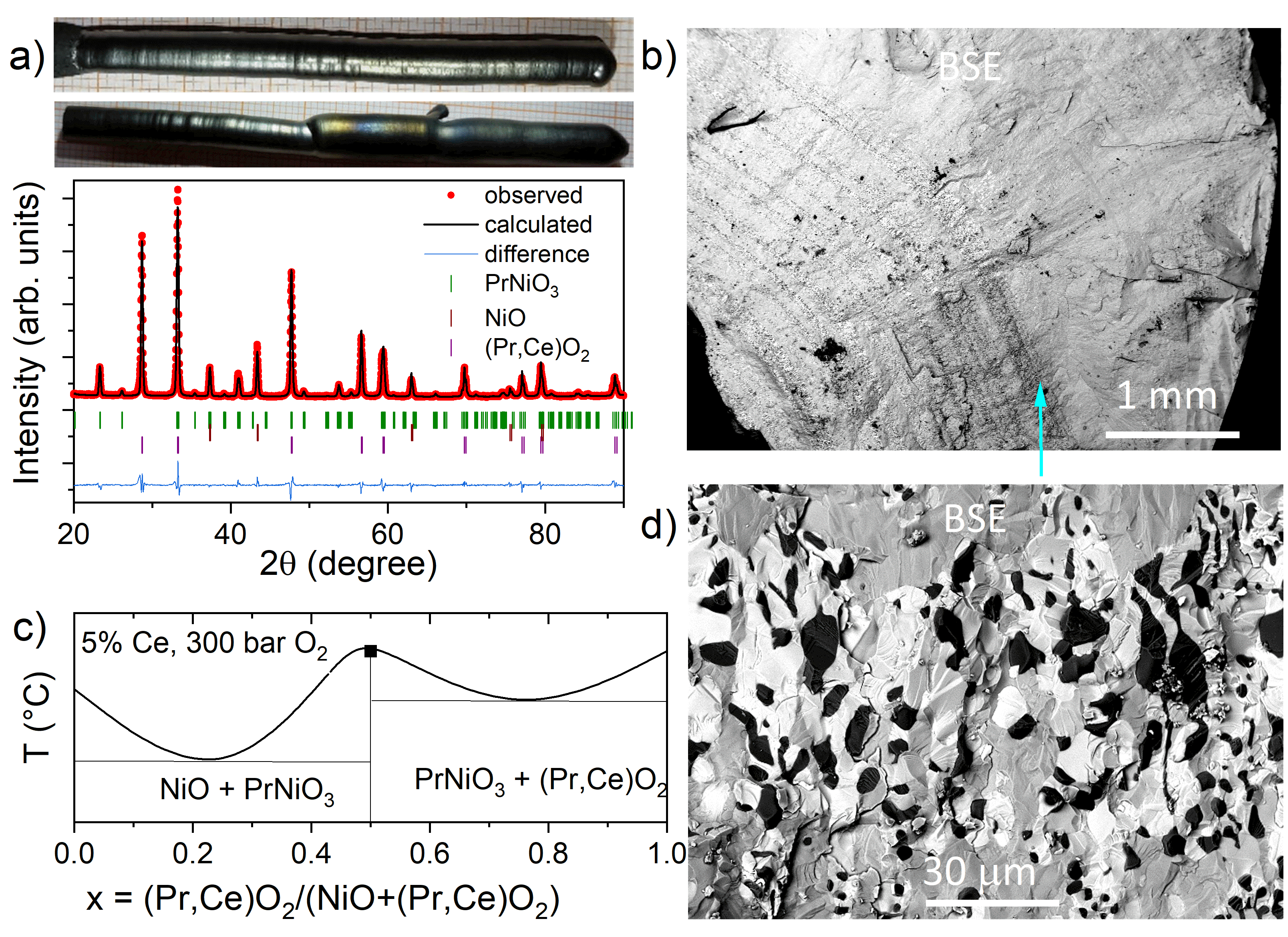}
\caption{(a) Images of the boules grown at 280 and 300 bar, and the PXRD pattern from a growth of the nominal composition Pr$_{0.95}$Ce$_{0.05}$NiO$_{3}$. The calculated Bragg peak positions of the Pr$_{1-x}$Ce$_{x}$NiO$_{3}$ majority (41.3(3) wt\%) and NiO (27.7(4) wt\%) and (Pr,Ce)O$_{2}$ (31.0(2) wt\%) minority phases are indicated as orange, purple, and green vertical bars, respectively. (b) BSE image of a broken surface from the grown boule, with the growth direction running from bottom left to top right. The blue arrow indicates lines of  eutectic freezing, which form when the solubility limit is reached. The regions with darker and lighter contrast correspond to NiO and (Pr,Ce)O$_{2}$, respectively. (c) Phase diagram for the phase formations of the OFZ melt under 300 bar of oxygen partial pressure for (Pr,Ce)O$_2$ and NiO. (d) Magnified BSE image of the same broken surface as in (b), but zoomed into an eutectic separation region.
}
\label{PCeNO}
\end{figure}

Electron-doping of PrNiO$_{3}$ can be achieved through substitution of the trivalent Pr ions by nominally tetravalent Ce ions. Here we carry out the OFZ growth at 280 and 300 bar oxygen partial pressure, respectively, aiming for 5\% Ce doping of the perovskite phase. The upper panel in Fig. \ref{PCeNO}a shows the as-grown boules for 280 (top) and 300 bar (bottom). In comparison to the growth of the undoped compounds, a more homogeneous and viscous liquid forms during this growth, facilitating the growth of a long and homogeneous boule. In our characterization, the boules grown under 280 and 300 bar show similar phase formations, and hence we only focus on the latter in the following.  Already for the small amount of 5\% doping, a phase mixture between Ce-substituted PrNiO$_{3}$, (Pr,Ce)O$_{2}$, and NiO forms (see PXRD in Fig. \ref{PCeNO}a), presumably because the solubility limit of Ce is very low. The phase formation in different regions of the boule is best seen in BSE imaging, with Fig. \ref{PCeNO}b giving a wide-scale overview of the cross section of the boule along the growth direction.

In the OFZ growth, we are constantly feeding the weighed-in stoichiometry, thus when a solubility limit in growth is reached the seed crystal incorporates too little of the corresponding element. Hence this accumulates in the growth until an eutectic point (the minimum in the phase diagram, e.g 0.2 and 0.75 shown in Fig. \ref{PCeNO}c) in the phase diagram is reached. Here, the eutectic part freezes out and forms a layer of the eutectic stoichiometry (which contains a major part of the impurity phase). This eutectic can be seen in Fig. \ref{PCeNO}b, which shows a large crossection of the boule via BSE with elemental resolution. Here black lines are visible revealing these eutectic rings. Figure \ref{PCeNO}d shows a magnified version of these eutectic mixes, revealing three phases intergrown with each other. Note that for all subsequent BSE images in the other subchapters we only show the magnified versions of the eutectic parts in our boule and XRD focuses on these sections as well, but in all cases similar eutectic rings are formed and completely homogenous parts exist as well, similar to what is visible in Fig. \ref{PCeNO}b on the top right. 

Nonetheless, the matrix is a perovskite phase as the Laue diffraction images from broken surfaces of the boule (not shown here) can be indexed in space group $Pbnm$, which suggests that large grains of Ce-substituted PrNiO$_{3}$ have crystallized in our boule, although BSE imaging reveals that the other two phases are intergrown in these grains (Figs. \ref{PCeNO}b,d). Importantly, we do not detect any traces of higher order Ruddlesden-Popper phases, such as Pr$_{4}$Ni$_{3}$O$_{10}$, which is in contrast to the Pr-excess growth of the undoped compound shown in Fig. \ref{PNO}e. Instead, we observe already the formation of (Pr,Ce)O$_{2}$ which is an endmember of the growth phase diagram. Hence, we propose the modified growth phase diagram shown in Fig. \ref{PCeNO}c for a Ce-mixed pseudo binary composition. 

\subsection{Electron-doping of LaNiO$_{3}$}

\begin{figure}[tb]
\includegraphics[width=1\columnwidth]{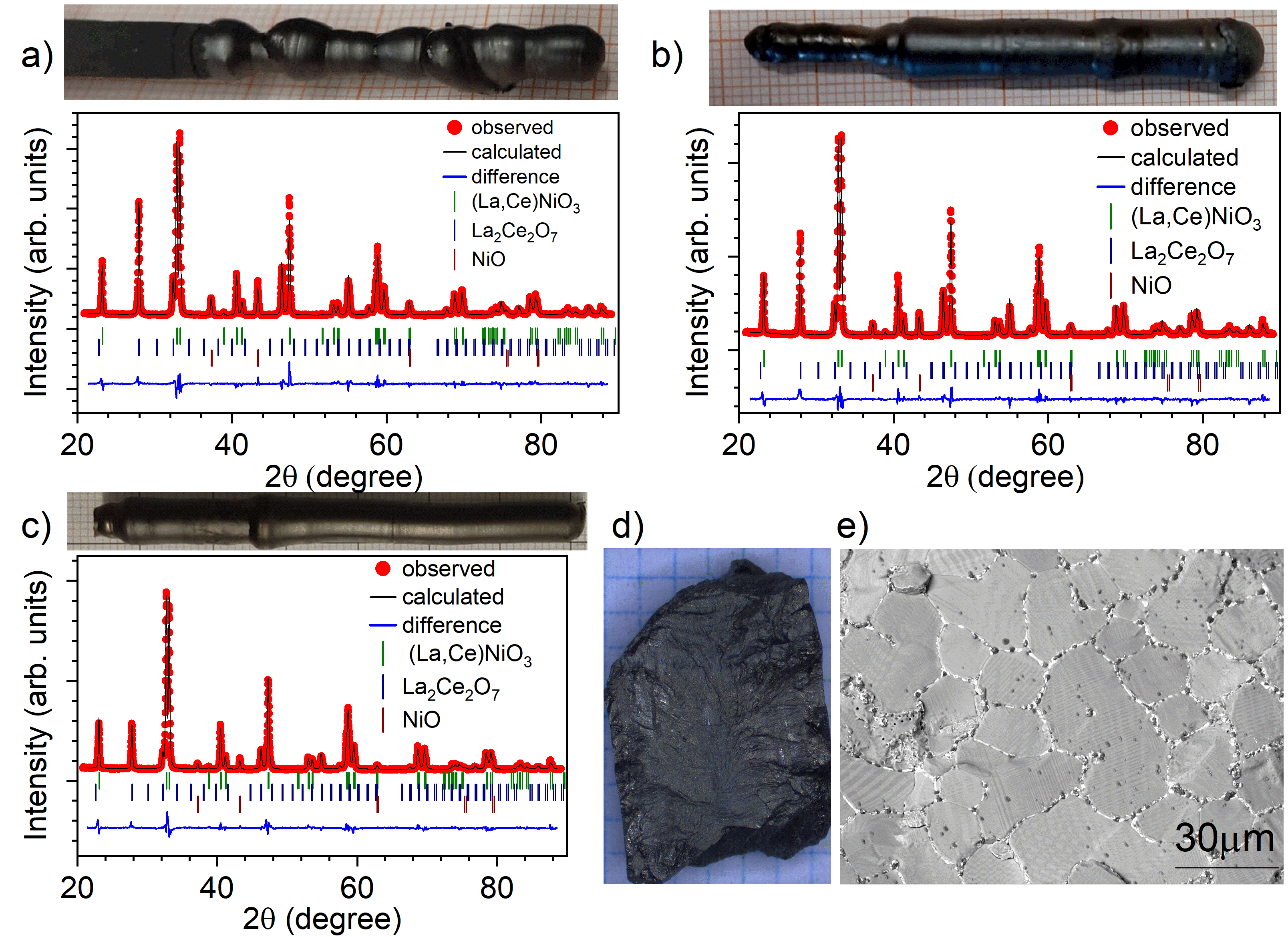}
\caption{Images of the boules and the PXRD patterns from growths of the nominal compositions (a) La$_{0.8}$Ce$_{0.2}$NiO$_{3}$, (b) La$_{0.95}$Ce$_{0.05}$NiO$_{3}$, and (c) La$_{0.91}$Ce$_{0.09}$NiO$_{3}$. The calculated Bragg peak positions of the La$_{1-x}$Ce$_{x}$NiO$_{3}$ majority (85.8(5) wt\%) and NiO (6.2(2) wt\%) and La$_{2}$Ce$_{2}$NiO$_{7}$ (8.00(7) wt\%) minority phases are indicated as green, red, and blue vertical bars, respectively. (d) Image of a cleaved single crystal of La$_{0.95}$Ce$_{0.05}$NiO$_{3}$, which was utilized in the XPS study. The growth direction runs from the bottom to the top. (e) BSE image of a surface of the La$_{0.91}$Ce$_{0.09}$NiO$_{3}$ boule cleaved perpendicular to the growth direction. Regions with darker and lighter contrast correspond to NiO and La$_{2}$Ce$_{2}$NiO$_{7}$, respectively.
}
\label{LCeNO}
\end{figure}

\begin{figure*}
\includegraphics[width=1.8\columnwidth]{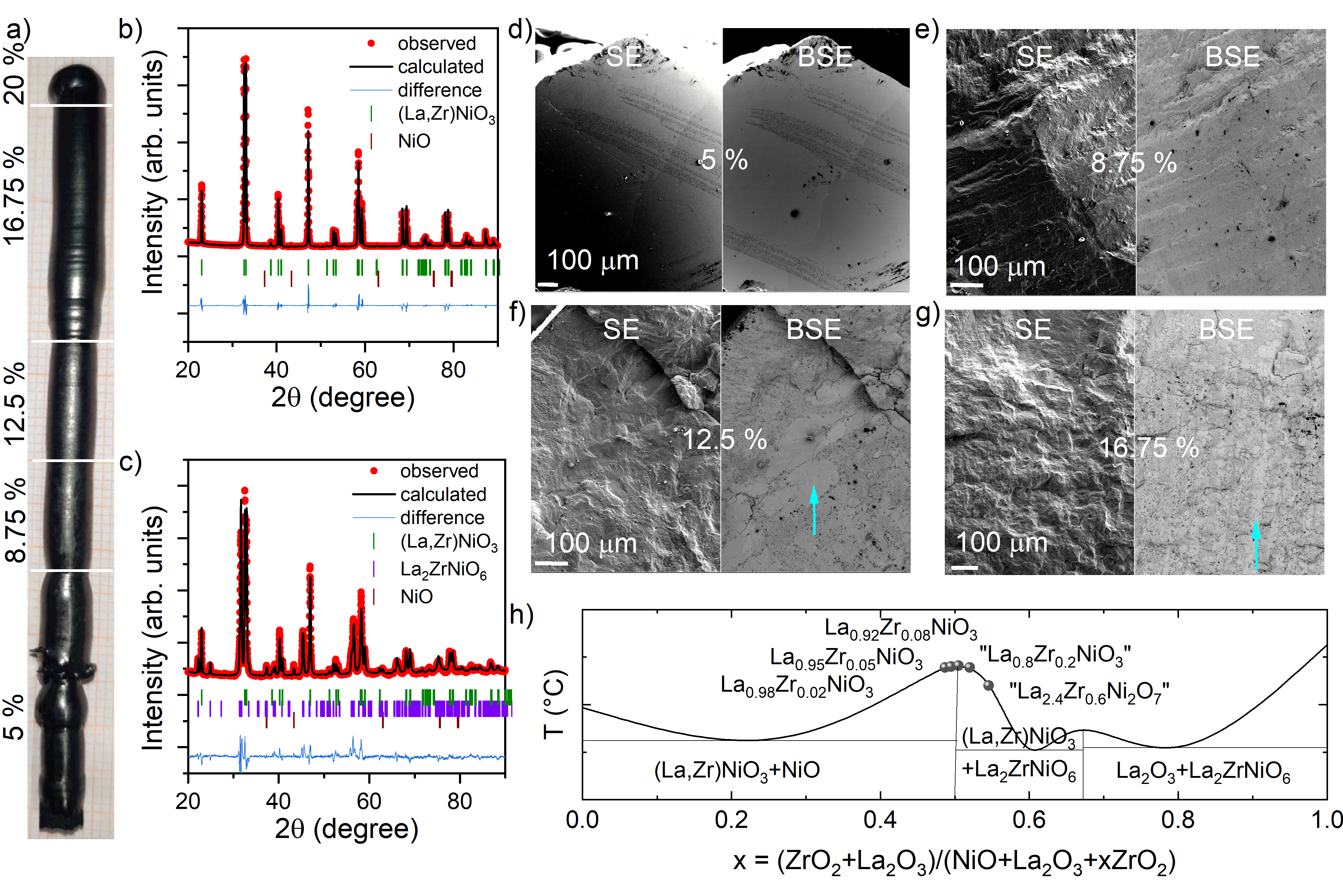}
\caption{(a) Image of the boule with a nominal La$_{1-x}$Zr$_{x}$NiO$_{3}$ composition, where the white lines separate sections with nominal Zr-substitutions of 5\%, 8.75\%, 12.5\%, 16.75\%, and 20\% (from bottom to top), respectively. (b,c) PXRD patterns of pulverized pieces from the sections with the nominal La$_{0.95}$Zr$_{0.05}$NiO$_{3}$ (b) and La$_{0.8}$Zr$_{0.2}$NiO$_{3}$ (c) compositions. The calculated Bragg positions of the La$_{1-x}$Zr$_{x}$NiO$_{3}$ majority and NiO minority phase are indicated by purple and green vertical bars, respectively. In (b) and (c), the majority phase contributes 99.5(2) wt\% and 57.8(5) wt\%, respectively, while the NiO minority phase corresponds to 0.5(2) wt\% and 2.8(2) wt\%, respectively. In addition, the La$_{2}$ZrNiO$_{6}$ double-perovskite phase is present (39.4(4) wt\%) in the PXRD in (c), as indicated by purple vertical bars. (d-g) SE (left) and BSE (right) images of broken surfaces from the sections with nominal Zr-substitutions of 5\% (d), 8.75\% (e) 12.5\% (f), and 16.75\% (g). The inclusions with a dark contrast in the BSE images correspond to NiO or dirt on the surface, while a lighter contrast (indicated by the arrows) in (f) and (g) corresponds to La$_{2}$ZrNiO$_{6}$ (highlighted by blue arrows). (h) Pseudo binary phase diagram of (Zr,La)O$_{x}$ and NiO at 85 bar oxygen partial pressure.
}
\label{LZNO}
\end{figure*}

Next, we focus on the electron-doping of LaNiO$_{3}$. As a first attempt we attempted to synthesize crystals of composition La$_{0.8}$Ce$_{0.2}$NiO$_{3}$, which according to prior work on infinite-layer thin films is optimal for superconductivity, but observed that the growth was not very stable as can be seen in the image of the grown boule in Fig. \ref{LCeNO}a. Large crystals could not be extracted from the growth and we find a low wt\% of the perovskite phase (see Tab. \ref{refinement}). Thus, we go for lower substitutions and we obtain large single-crystals with millimeter dimensions (as shown in Fig. \ref{LCeNO}a) for the nominal compositions La$_{0.95}$Ce$_{0.05}$NiO$_{3}$ and La$_{0.91}$Ce$_{0.09}$NiO$_{3}$. For the La-based compounds, we observe that already small Ce-substitution levels lead to the emergence of a competing phase, which in this case is the pyrochlore phase La$_{2}$Ce$_{2}$O$_{7}$. The presence of this secondary phase is detected in the PXRD patterns (Figs. \ref{LCeNO}b,c) and the BSE image (Fig. \ref{LCeNO}e), revealing that  La$_{2}$Ce$_{2}$O$_{7}$ is embedded in the single crystalline perovskite matrix as eutectic rings together with NiO. Hence, Ce-substitution of the perovskite phase is challenging, although small crystals of (La,Pr)$_{1-x}$Ce$_{x}$NiO$_{3}$ surrounded by regions of secondary phases can be realized in the boule.

A different route to achieve electron-doping can be the substitution by tetravalent Zr ions. Such doping was recently proposed theoretically for the topotactically reduced $n =2 $ compound La$_{2.4}$Zr$_{0.6}$Ni$_{2}$O$_{6}$ \cite{Worm2022} where superconducting behavior is expected according to the nominal valence of Ni. As a first step, we attempt to synthesize this material at a moderate pressure of 14 bar oxygen. While such pressure was sufficient to synthesize the non-substituted $n =2 $ compound La$_{3}$Ni$_{2}$O$_{7}$ \cite{Liu2022}, we find for the Zr-doped variant that a phase mixture between the perovskite and the double-perovskite phase La$_{2}$NiZrO$_{6}$ forms (see Tab. \ref{refinement}). As the $n = \infty$ phase is typically stable starting with 30 bar \cite{Zhang2017LNO}, we reduce the pressure and attempt a synthesis at 2.5-4.5 bar (see Tab. \ref{refinement}). In this case, however, we obtain a mixture of the $n=0$ and $n = \infty$ phases, finally proving the destabilization of higher order Ruddlesden-Popper phases with electron doping. 

As a next step, we focus on the synthesis of the perovskite phase and test the maximally possible Zr substitution content. To this end, we prepare a ''gradient rod'' that is composed of sections with different substitution levels of Zr. The prepared sections correspond to 5\%, 8.75\%, 12.5\%, 16.75\%, and 20\% Zr (Fig. \ref{LZNO}a). During the growth, we observe that all substitutions help to stabilize the melt, while secondary phases emerge for 8.75\% and higher substitutions. Notably, the growth conditions are stable even in the case of evolving phase mixtures. However, no large single-crystals form in these attempts due to the shrinking sizes of the perovskite domains, as evidenced in the BSE images, seen in the decrease of a single contrast matrix (see Figs. \ref{LZNO}f, g). A rudimentary proposal for the pseudo-binary phase diagram is shown in Fig. \ref{LZNO}h, which notably only provides points in the center, where our PXRD characterization (exemplary PXRD patterns in Figs. \ref{LZNO}b,c) reveal the formation of the pure perovskite and  perovskite double-perovskite mixtures at higher doping levels, respectively. 

Additionally, we explore the phase formation for the growth of La$_{0.8}$Zr$_{0.2}$NiO$_{3}$ at various oxygen partial pressures ranging from 15, 25, 35 to 150 bar (not shown here). We find that for a stable growth of the perovskite phase at least 35 bar pressure is required. Above this pressure, however, we do not detect significant changes in the phase formation, nor the incorporated Zr-substitution content. These results indicate that the solubility limit of Zr in nickelates lies around 8\%, which is also confirmed by a detailed EDS analysis (see Appendix B), where even for crystals from a growth with a nominally higher Zr substitution the detected content in the perovskite matrix does not exceed 8\%. The formed double-perovskite phase can be considered as an impurity phase where Zr substitutes half of the Ni site, instead of the desired substitution of the La sites. Hence, one might suspect that Zr generally substitutes the Ni site, even for low Zr admixtures. However, our Rietveld refinement of the perovskite phase reveals that in this phase Zr solely occupies the La site. Moreover, we observe a solubility limit of approximately 8\% also for substitutions with elements other than Zr (see below).

\begin{figure}[tb]
\includegraphics[width=1\columnwidth]{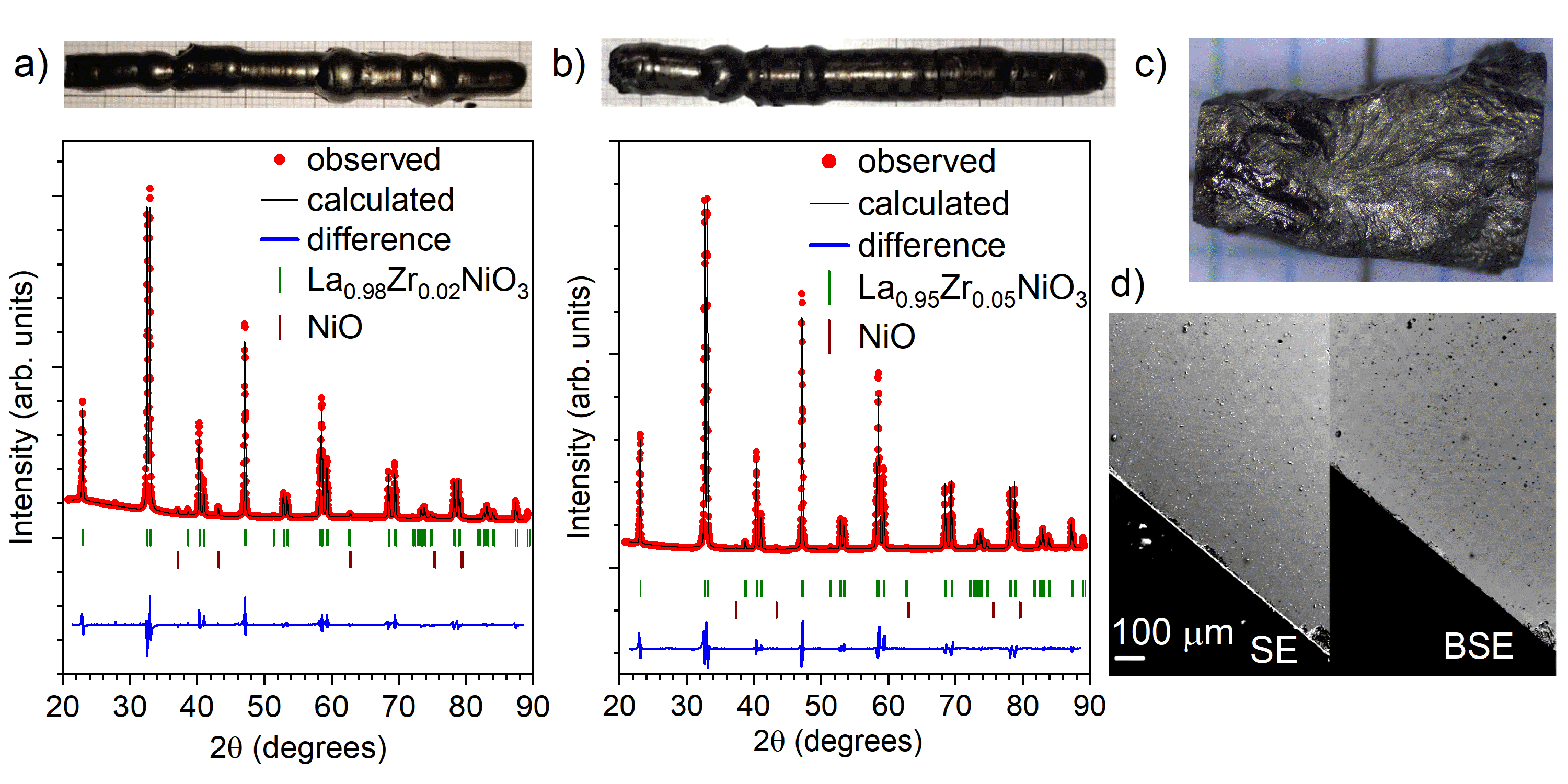}
\caption{(a,b) Images of the boules and the PXRD patterns from the growths of the nominal compositions La$_{0.98}$Zr$_{0.02}$NiO$_{3}$ (a) and La$_{0.95}$Zr$_{0.05}$NiO$_{3}$ (b). The calculated Bragg peak positions of the majority La$_{1-x}$Zr$_{x}$NiO$_{3}$ (97.5(4) wt\% and 99.5(6) wt\%, respectively) and minority NiO phases (3.5(5) wt\% and 0.5(2) wt\%, respectively) are indicated by green and brown vertical bars, respectively. (c) Image of a cleaved single crystal from the La$_{0.95}$Zr$_{0.05}$NiO$_{3}$ boule. (d) SE (left) and BSE (right) images of the polished surface of the La$_{0.95}$Zr$_{0.05}$NiO$_{3}$ crystal analyzed in Fig. \ref{res}. The dark spots correspond to NiO.}
\label{LeNO}
\end{figure}

Motivated by these insights, we next carry out the growth of crystals with much lower Zr substitution content and optimized growth parameters. Pictures of the corresponding La$_{0.98}$Zr$_{0.02}$NiO$_{3}$ and La$_{0.95}$Zr$_{0.05}$NiO$_{3}$ boules are shown in  Fig. \ref{LeNO}. From the boules, we extract large and high-quality perovskite crystals, as confirmed by PXRD (Figs. \ref{LeNO}a,b) and BSE imaging (Fig. \ref{LeNO}d), as well as EDS analysis (see Appendix B).

\subsection{Hole-doping of LaNiO$_{3}$}

Hole-doping of perovskite nickelates can be achieved through the substitution with divalent ions such as Ca and Sr. In a previous high-pressure flux growth, the synthesis of small Ca-substituted LaNiO$_{3}$ single-crystals was achieved \cite{Puphal2021}, with 8\% Ca content determined by XRD refinement, whereas EDS indicated a higher level of 12\% in cleaved pieces. In an attempt to realize larger Ca-substituted crystals with the floating zone technique under oxygen pressures ranging from 200-300 bar, we perform growths with nominal compositions La$_{0.8}$Ca$_{0.2}$NiO$_{3}$ and Pr$_{0.8}$Ca$_{0.2}$NiO$_{3}$. However, we find that only the $n = 1$ Ruddlesden-Popper phases La$_{1.6}$Ca$_{0.4}$NiO$_{4}$ and Pr$_{1.6}$Ca$_{0.4}$NiO$_{4}$ as well as NiO form for such high substitution levels (not shown here). Since our floating zone growth with Zr revealed a solubility limit around 8\% (Fig. \ref{LZNO}e), we next attempt the growth of La$_{0.92}$Sr$_{0.08}$NiO$_{3}$. However, we find that connecting the corresponding feed and seed rods at 300 bar pressure is unfeasible, due to a heavy oxidization and freezing of the melt. To circumvent these issues we start the growth at a lower pressure and by sequentially increasing it up to 300 bar a stable growth is achieved. The results of the growth are presented in Tab. \ref{refinement}.

\begin{figure}[tb]
\includegraphics[width=1\columnwidth]{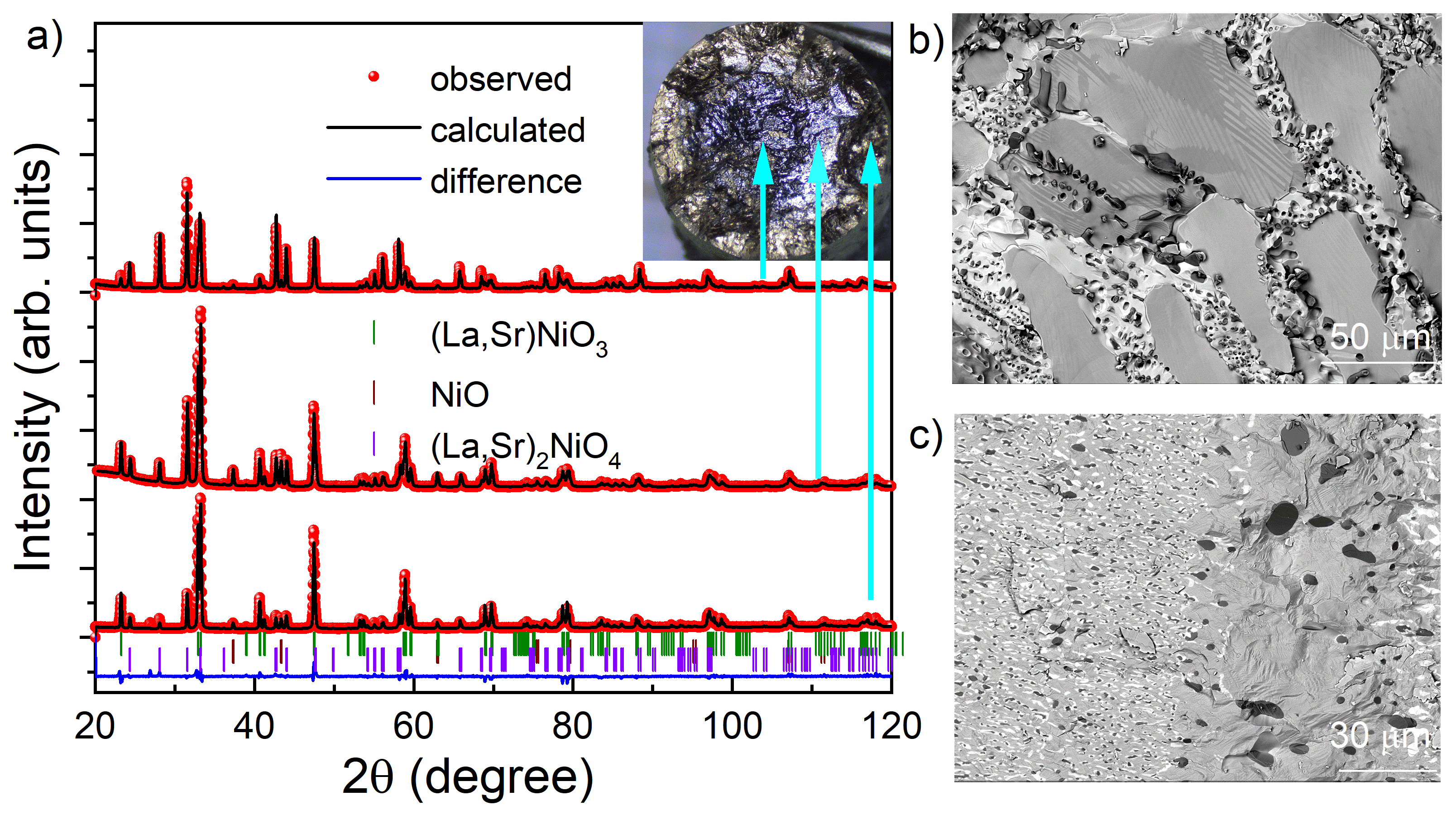}
\caption{Image of the cross section of the boule (inset) and PXRD patterns from a growth of the nominal composition La$_{0.84}$Sr$_{0.16}$NiO$_{3}$. The three PXRD patterns are taken from pieces from different positions in the boule, as indicated by the blue arrows. The calculated Bragg peak positions (vertical bars) and the difference between the experimental and calculated (blue line) are shown for the bottom PXRD pattern. The corresponding wt\% of the phases are summarized in Tab. \ref{refinement}.
(b,c) BSE images from different parts of the cross section revealing a perovskite matrix with NiO (dark) and  (La,Sr)$_{2}$NiO$_{4}$ (light) eutectic inclusions.
}
\label{LSNO}
\end{figure}

To confirm the anticipated the solubility limit of Sr of 8\%, we subsequently attempt the growth of La$_{0.84}$Sr$_{0.16}$NiO$_{3}$. Unlike in the 8\% case, a direct connection at 300 bar pressure is possible, highlighting that an increase in pressure is necessary to stabilize the higher nominal Ni$^{3.16+}$ oxidation state compared to Ni$^{3.08+}$ in La$_{0.92}$Sr$_{0.08}$NiO$_{3}$. Figure \ref{LSNO}a shows the PXRD of pulverized pieces broken off from different parts of the boule. As expected for the growth of hole-doped compounds where extreme oxygen pressures are required, a concentration gradient is observed towards the center of the boule (see inset in Fig. \ref{LSNO}a). In particular, we find that the fraction of the desired perovskite phase decreases towards the center of the boule, which is in accord with an increasing oxygen deficiency reported in previous floating zone growths of LaNiO$_{3}$ \cite{Zheng2020}. The wt\% and lattice constants from three different regions of the boule are summarized in Tab. \ref{refinement}. 

As a secondary phase in the PXRD, we detect the $n = 1$ Ruddlesden-Popper phase, which is also evidenced in the BSE imaging as a light color (Figs. \ref{LSNO}b,c). As can be seen especially in Figs. \ref{LSNO}b,c,  in the center of the boule the grains of the perovskite phase are immersed in a matrix of the $n = 1$ Ruddlesden-Popper phase and NiO (Fig. \ref{LSNO}b), whereas the situation is reversed in the outer regions, where the Ruddlesden-Popper phase and NiO form small inclusions in the perovskite phase (Fig. \ref{LSNO}c). Such an intergrowth behavior appears to be a general characteristic for all investigated substitutions above 8\% and points towards the existence of a general solubility limit in perovskite nickelates independent of the dopant species. In contrast, substitution levels far above 8\% have been reported for thin film perovskite nickelates \cite{Lee2020}, which might be facilitated by the epitaxial strain provided by the substrate. 

Nevertheless, our EDS analysis (see Appendix B) performed locally on grains with the perovskite phase indicates that the Zr and Sr contents can reach values between 12 and 13\% (see Tab. \ref{refinement}), which is comparable to the highest value detected by EDS in high-pressure flux grown Ca-substituted LaNiO$_{3}$ single-crystals \cite{Puphal2021}. These values are substantially higher 8\%, but it is possible that the EDS analysis includes a systematic overestimation of these dopants. In contrast, for the rare-earth element Ce, where the EDS analysis is based on the $M$ instead of the $K$-line, we find a maximum of 7(1)\% substitution (Tab. \ref{refinement}), which is compatible with a solubility limit of 8\%. In conclusion, complementary studies with analyses methods other than EDS are desirable to determine the maximum substitution content of Zr and Sr realizable in optical floating zone grown perovskite nickelates.

\subsection{Physical properties}

Recent investigations into the effects of hole- and electron-doping on perovskite nickelates have primarily utilized thin film samples \cite{Song2023,Patel2022}. In electron-doped Nd$_{1-x}$Ce$_{x}$NiO$_{3}$, films, a suppression of the metal-to-insulator transition of NdNiO$_{3}$ was reported, favoring a shift towards a metal-to-metal transition for Ce doping levels as subtle as 2.5\% \cite{Song2023}. In the following, we explore similar effects in our (Pr,Ce)NiO$_{3}$ crystals.

\begin{figure}[tb]
\includegraphics[width=1\columnwidth]{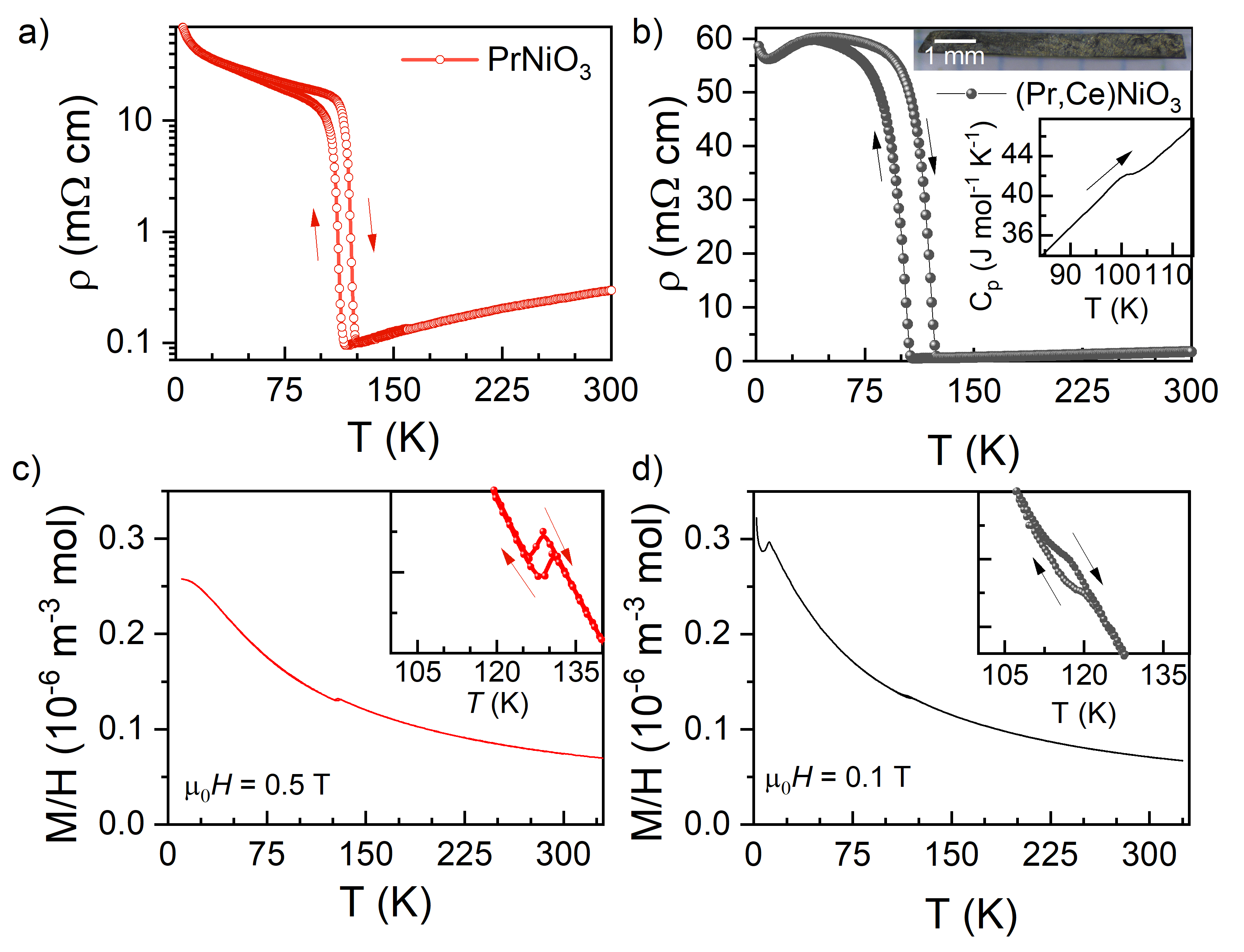}
\caption{(a,b) Resistivity as a function of temperature for a  PrNiO$_{3}$ (a) and a Pr$_{0.95}$Ce$_{0.05}$NiO$_{3}$ crystal (b). The current direction coincides with the $c$-axis direction. The upper inset in (b) shows the cut Pr$_{0.95}$Ce$_{0.05}$NiO$_{3}$ crystal and the lower inset the specific heat measured upon cooling. (c,d) Magnetization as a function of temperature for the PrNiO$_{3}$ (c) and Pr$_{0.95}$Ce$_{0.05}$NiO$_{3}$ crystal (d). The insets display a magnification of the hysteresis loops observed around the magnetic transitions. 
}
\label{resP}
\end{figure}

Figure \ref{resP} presents a comparative analysis of the electronic and magnetic properties of our undoped and Ce-doped PrNiO$_{3}$ crystals. Note that the crystals have been cut into oriented pieces and annealed in a high gas pressure autoclave at 600$^{\circ}$C under an oxygen partial pressure of 600 bar for 5 days, followed by a rapid cooling down to room temperature. This methodical approach ensures the exclusion of the influences of oxygen deficiency on the measured properties. Figure \ref{resP}a displays the expected sharp metal-to-insulator transition over several orders of magnitude in the resistivity in the undoped PrNiO$_{3}$ single-crystal around 128 K, consistent with previous literature \cite{Zheng2019}. For our single crystals with a nominal composition of Pr$_{0.95}$Ce$_{0.05}$NiO$_{3}$, EDS indicates a lower realized Ce substitution level of around 2(1)\%. We observe a metal-to-metal transition around 116 K in Fig. \ref{resP}b. This change from a metal-to-insulator to a metal-to-metal transition is qualitatively similar to that in Nd$_{1-x}$Ce$_{x}$NiO$_{3}$ films \cite{Song2023}. 

In the magnetic susceptibility, anomalies with hysteretic behavior occur at similar temperatures as the metal-to-insulator and metal-to-metal transition in PrNiO$_{3}$ and (Pr,Ce)NiO$_{3}$, respectively (Figs. \ref{resP}c,d). These anomalies indicate the presence of an antiferromagnetic transition \cite{Zheng2019}. Above the transition, the susceptibility can be well described by a Curie-Weiß law $\chi$ = $\frac{C}{T - \theta_W}$, which yields the Curie constant $C=2.20(6)$ emu K/mol Oe$^{-1}=27.6(7)\cdot10^{-6}$m$^3$K/mol and $\theta_W=-91(1)$ K for (Pr,Ce)NiO$_{3}$. This is well comparable to undoped PrNiO$_3$ with a Curie constant $C=2.38(1)$ emu K/mol Oe$^{-1}=29.9(1)\cdot10^{-6}$m$^3$K/mol and $\theta_W=-99(1)$ K. We note that the magnetic signal of our (Pr,Ce)NiO$_{3}$ sample contains a contribution from the intergrown (Pr,Ce)O$_{2}$ phase, which exhibits an antiferromagnetic transition at 12 K whose signature can be seen both in the resistivity and susceptibility (Figs. \ref{resP}b,d). 

The susceptibility curves in Figs. \ref{resP}c,d are expected to be dominated by the paramagnetic signal of the Pr$^3+$ cations. Nevertheless, the effective moment is extracted using $\mu_{eff}=\sqrt{3kC/N\mu_B^2}=2.828\sqrt{C}$. Assuming the effective paramagnetic moment of non-disproportionated Ni$^{3+}$ LS ($\mu_{eff} = 1.73 \mu_B=\sqrt{3kC/N\mu_B^2}$), $C_{Ni^{3+}}$ is estimated to be $0.375$ emu K/mol. Subsequently, if we consider $C=C_{Ni}+C_{Pr}$, $\mu_{eff}$ of Pr is calculated to be 3.82 $\mu_B$ which is only slightly larger than the expected moment of 3.5 $\mu_B$ and the reported one for PrNiO$_3$ \cite{Klein2021}.

\begin{figure}[tb]
\includegraphics[width=1\columnwidth]{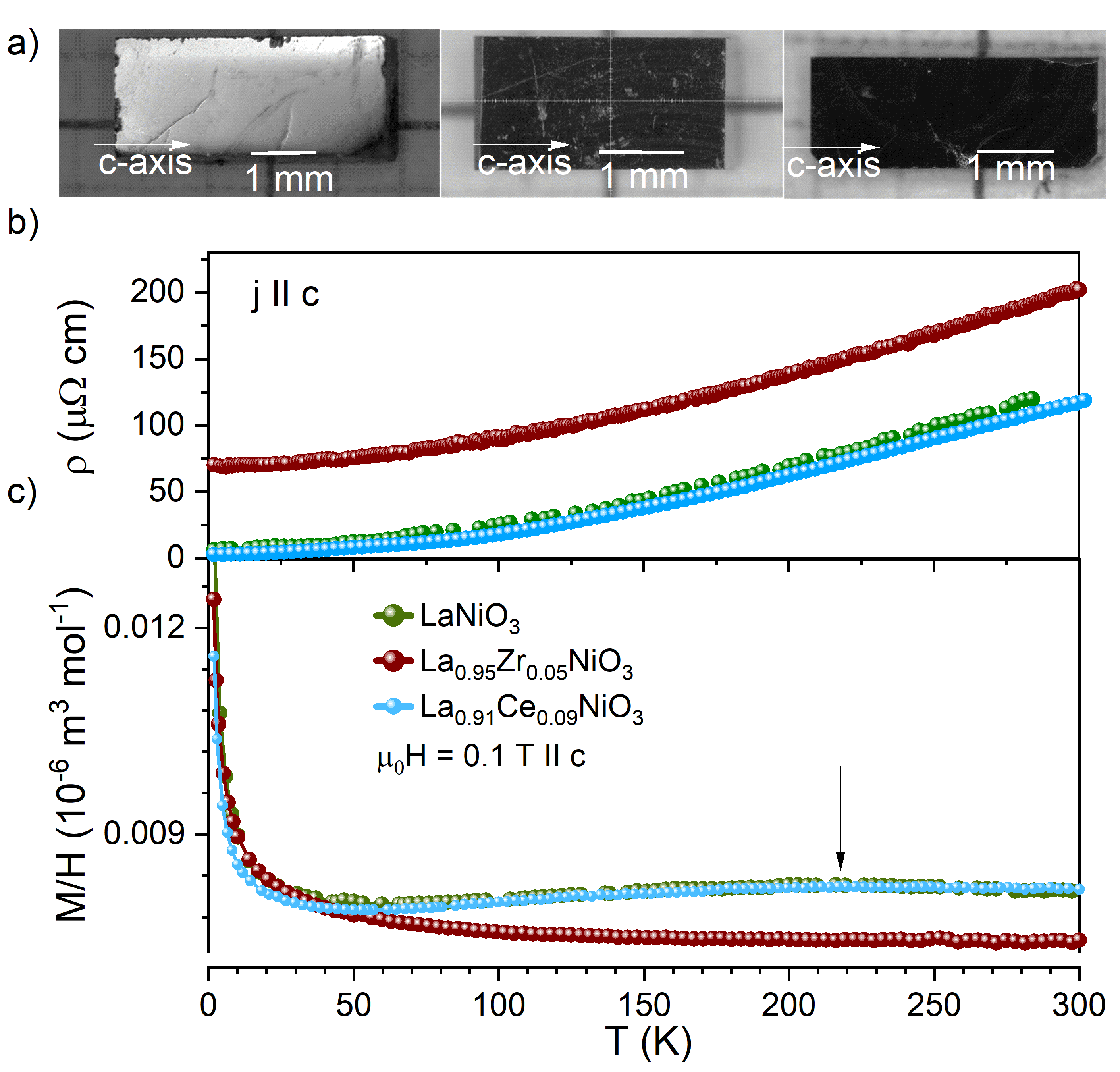}
\caption{(a) Images of the shaped and polished LaNiO$_{3}$, La$_{0.91}$Ce$_{0.09}$NiO$_{3}$, and La$_{0.95}$Zr$_{0.05}$NiO$_{3}$ single-crystals (left to right), with horizontal $c$-axis orientation. (b) Resistivity as a function of temperature for the three crystals. The current direction coincides with the $c$-axis direction. (c) Magnetization  as a function of temperature for the three crystals. A field of 0.1 T was applied along the $c$-axis direction. 
}
\label{res}
\end{figure}

Next, we characterize the physical properties of a selection of La-based compounds. To this end, we analyze large crystals of LaNiO$_{3}$, La$_{0.95}$Zr$_{0.05}$NiO$_{3}$, and La$_{0.91}$Ce$_{0.09}$NiO$_{3}$ by Laue diffraction and cut them into rectangular pieces with the $(1\bar10)$ direction pointing out of the plane, while the $c$-axis and (110) direction lie in the horizontal plane (Fig. \ref{res}a). Subsequently, the surfaces of the shaped and oriented crystals are polished using a 1 $\mu$m grain size. The crystals are then annealed in a high gas pressure autoclave at 600$^{\circ}$C under an oxygen partial pressure of 600 bar for 5 days, followed by quick cooling to room temperature. We find that oxygen annealing is required to alleviate the oxygen deficiency of the as-grown crystals, which induces antiferromagnetic order\cite{Zeng2020}.
Figure \ref{res}b) shows the resistivity of the annealed crystals measured along the $c$ axis. The high quality of our single-crystals is reflected in the large residual resistivity ratio (RRR) of LaNiO$_3$, with a value of 19, exceeding previously reported values \cite{Zhang2017LNO,Zheng2020,Dey2019,Tomioka2021,Puphal2023}. The RRR value of the Zr-substituted crystal is found to be 3, and that of the Ce-substituted crystal is 43. The slope of the resistivity curve of all three crystals is similar, with Fermi-liquid behavior at temperatures below 100 K and a $T^{1.5}$ scaling behavior at higher temperatures. This is consistent with earlier reports of OFZ grown perovskite nickelates \cite{Zhang2017LNO, Guo2018}. While the Ce substitution lowers the overall resistivity, as expected for electron doping, Zr substitution appears to show the opposite effect. 

Due to the metallic nature, the magnetic properties of LaNiO$_3$ are not as simple as following the expected Curie laws, unlike PrNiO$_3$. Pure LaNiO$_3$ has complex magnetic correlations that can not be described by simple Pauli paramagnetism \cite{Wang2018}. A Stoner enhanced maximum (highlighted by an arrow) is observed in LaNiO$_3$ centered around 220 K (Fig. \ref{res}c), which cannot by described by simple Pauli paramagnetism. Zr doping leads to a complete suppression of this observed maximum in the magnetic susceptibility. As a result, the full temperature range of the susceptibility of the Zr doped compound can be captured by a Curie-Weiß fit, with $C=1.504(3)\cdot 10^{-3}$ emu K/mol Oe$^{-1}=1.89(4) \cdot 10^{-8}$ m$^3$K/mol and $\theta_W=-2.2(1)$ K. As discussed above, we take $C_{Zr}$ to be $C-0.375$. This results in $\mu_{eff}=3\mu_B$ which is comparable to an expected moment $\mu_{Zr^{2+}}=2.8\mu_B$ of a $J=2$ state, suggesting a magnetic Zr contribution. Notably, on the other hand we observe a similar Stoner enhanced maximum and a somewhat lower Curie tail for Ce substitution. In  La$_{0.91}$Ce$_{0.09}$NiO$_{3}$, however,  a considerable mass fraction is La$_2$Ce$_2$O$_7$. La$_2$Ce$_2$O$_7$ is solely diamagnetic and results in a small contribution in the susceptibility. Thus for the Ce doped crystal, the real mass of the perovskite phase was considered, determined from PXRD via Rietveld refinement.

\begin{figure*}[tb]
\includegraphics[width=2.0\columnwidth]{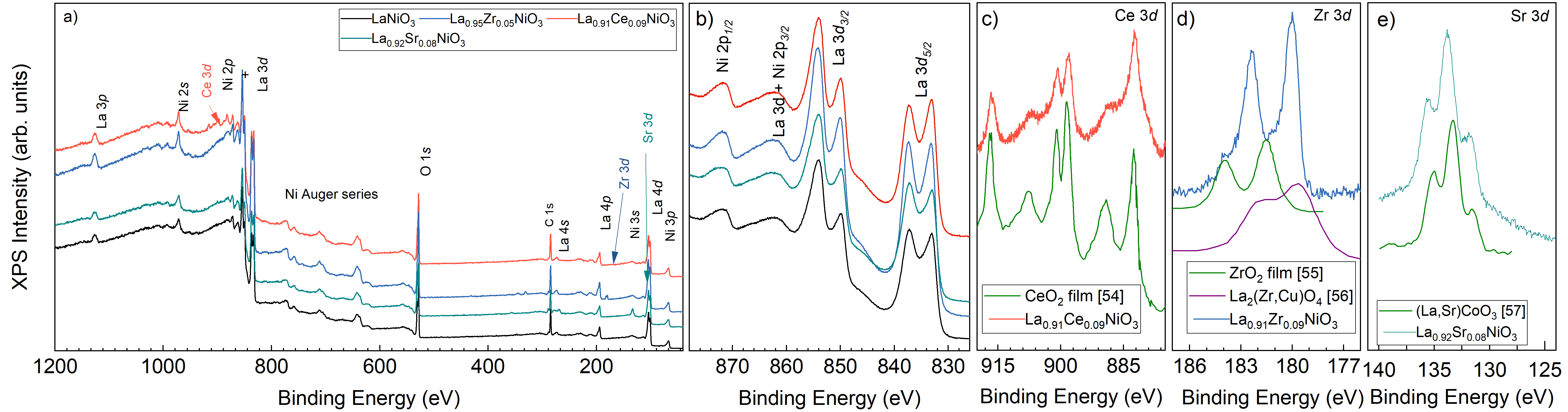}
\caption{(a) XPS spectra of LaNiO$_{3}$ (black) \cite{Puphal2023}, La$_{0.92}$Sr$_{0.08}$NiO$_{3}$ (green), La$_{0.95}$Zr$_{0.05}$NiO$_{3}$ (blue), and La$_{0.91}$Ce$_{0.09}$NiO$_{3}$ (red) across a wide binding energy range from 100-1200 eV. (b) Zoom into the region between 875 and 825 eV, which includes the La 3$d$ and Ni 2$p$ edges. (c) Zoom into the region between 916.5 and 881.5 eV, which includes the Ce 3$d$ edge. Data of a CeO$_2$ film from Ref.~\cite{Pfau1996} are shown for comparison. (d) Zoom into the region between 187 and 176 eV, corresponding to the Zr 3$d$ edge. Data of a ZrO$_2$ film from Ref.~\cite{Lackner2019} and a La$_2$(Zr,Cu)O$_4$ pellet  from Ref.~\cite{Anderson1994} are shown for comparison. (e) Zoom into the region between 140 and 125 eV, corresponding to the Sr 3$d$ edge. Data of a (La,Sr)CoO$_3$ from Ref.~\cite{Wang2018X} are shown for comparison. 
}
\label{XPS}
\end{figure*}

\subsection{Valence states}

To further characterize our samples and determine the oxidation states of the substituted ions, we investigate freshly cleaved surfaces of selected single crystals using XPS. Specifically, this technique offers insights into the effectiveness of the intended charge carrier doping. Due to the significant overlap and complex multi-component structures of the La 3$d$ and Ni 2$p$ core-level spectra, we will concentrate on a qualitative analysis in the following. 

An overview of all XPS spectra across a wide binding energy range is displayed in Fig. \ref{XPS}a. In Fig. \ref{XPS}b, the peaks at 854.8 and 851.2 eV correspond to La 3$d_{3/2}$, while the doublet at 837.9 and 834.6 eV can be identified as La 3$d_{5/2}$, indicating the La$^{3+}$ oxidation state. The La 3$d_{3/2}$ and Ni 2$p_{1/2}$ peaks overlap \cite{Mickevicius2006}, and the Ni 2$p_{3/2}$ peak is accompanied by a satellite line, which is positioned approximately 6–7 eV higher in binding energies. The Ni 2$p_{1/2}$ peak follows at 872.3 eV.

Figure \ref{XPS}c focuses on the XPS of La$_{0.91}$Ce$_{0.09}$NiO$_{3}$. Ce 3$d$ is known to exhibit a very complex multiplet splitting, resulting in several peaks. Notably, the multiplet structure observed in our La$_{0.91}$Ce$_{0.09}$NiO$_{3}$ spectrum matches the characteristic spectrum of  Ce$^{4+}$ in CeO$_2$ \cite{Pfau1996}, thereby conﬁrming the presence of carrier doping in this system since Ce$^{4+}$ is replacing La$^{3+}$ sites. In contrast, Ce$^{3+}$ would give rise to two doublets at 880.9 eV / 885.2 eV and 899.1 eV / 903.4 eV \cite{Beche2008}, which we do not observe. 

In the case of the presence of Zr$^{4+}$ in electron-doped La$_{1-x}$Zr$_{x}$NiO$_{3}$, the corresponding peak of the Zr 3$d_{5/2}$ binding energy is expected around 182 eV \cite{Lackner2019}. However, we observe two Zr  3d$_{5/2}$ doublets, corresponding to binding energies of 181.1 eV and 180 eV, clearly indicating a lower Zr valency, and likely even a mixed valence state (see reference data from ZrO$_{2}$ in Fig. \ref{XPS}d). This suggests that the anticipated electron doping might not have occurred in this system, which is  consistent with the increase in resistivity observed in the Zr-substituted crystal, as opposed to the decrease in resistivity in the Ce-substituted crystal (Fig. \ref{res}b). Nevertheless, we cannot rule out that the shape of the spectrum in Fig. \ref{XPS}d is strongly influenced by surface effects, as in the case of Sr-doping, which is described below.

Lastly, we explore the possibility of achieving hole-doping by incorporating Sr$^{2+}$ ions in La$_{0.92}$Sr$_{0.08}$NiO$_{3}$. Previous works 
have suggested that hole doping is realized in Sr-substituted nickelate thin films \cite{Li2019,Zeng2020,Lee2020,Gao2021,Osada2021}, but no direct evidence of the Sr$^{2+}$ valence state has been reported so far, to the best of our knowledge. In the XPS of our La$_{0.92}$Sr$_{0.08}$NiO$_{3}$ crystal, we find that the Sr 3$d$ multiplet exhibits several overlapping peaks (Fig. \ref{XPS}e), in particular the 3$d_{5/2}$ peaks at 131.8 eV and 133.8 eV, and the 3$d_{3/2}$ peaks at 133.6 eV and 135.6 eV. These peaks differ from the simpler 3$d$ doublet  at 134.3 eV and 132.5 eV of Sr$^{2+}$ in SrTiO$_{3}$ \cite{Vasquez1992}. Yet, our observed peak structure is closely similar to that of the Sr ions in La$_{0.4}$Sr$_{0.6}$CoO$_3$\cite{Wang2018X}. In the cobaltate, the two doublets were ascribed to distinct surface and lattice Sr sites, due to the formation of SrO and Sr(OH)$_2$ on the surface. Such a scenario could apply similarly to the Sr ions in our nickelate sample.


\section{\label{sec:level4}Summary}
In summary, we have successfully grown single-crystals of LaNiO$_{3}$, PrNiO$_{3}$, La$_{0.98}$Zr$_{0.02}$NiO$_{3}$, La$_{0.95}$Zr$_{0.05}$NiO$_{3}$, and La$_{0.92}$Sr$_{0.08}$NiO$_{3}$. On the other hand, the (La,Pr,Eu)$_{6}$Ni$_{5}$O$_{16}$ and La$_{2.4}$Zr$_{0.6}$Ni$_{2}$O$_{7}$ phases did not form under the employed growth conditions. The synthesis of Pr$_{0.95}$Ce$_{0.05}$NiO$_{3}$, La$_{0.95}$Ce$_{0.05}$NiO$_{3}$, La$_{0.91}$Ce$_{0.09}$NiO$_{3}$, and La$_{0.84}$Sr$_{0.16}$NiO$_{3}$ was accomplished, albeit with the drawback of substantial amounts of impurity inclusions in the matrix. In general, we find that substitutions higher than 8\%, or any electron doping of the Ruddlesden-Poppers phases $n=2,3$ lead to phase separation for OFZ growth under 300 bar oxygen pressure.

Our physical properties and XPS characterizations reveal that electron-doping is realized for Ce-substituted PrNiO$_{3}$ and LaNiO$_{3}$, whereas Zr-substitution likely results in mixed-valent states, potentially affecting the metallic ground state and the Stoner enhanced maximum in the magnetic susceptibility. 

Our results underscore the pressing need for new, cutting-edge strategies that transcend the conventional single dopant route to realize bulk crystals of nickelate superconductors. Thus, co-doping with more than one ionic species or self doping via the oxygen stoichiometry, as previously established in cuprates\cite{Frano2019}, might offer a path to realize superconductivity in topotactically reduced nickelate crystals.

\begin{acknowledgments}
We thank F. Predel for acquiring BSE images at an early stage of this project, and C. Busch for technical support. We acknowledge helpful discussions with S. Hayashida and are grateful for access to the Merlin SEM at the Scientific Facility Nanostructuring Lab (NSL) at MPI FKF.
\end{acknowledgments}

\appendix
\section{Pressure dependence}

Before the OFZ growth of La$_{1-x}$Zr$_{x}$NiO$_{3}$ substitution series, we performed a study on the growth pressure during a single growth of La$_{0.8}$Zr$_{0.2}$NiO$_{3}$. Our initial attempt employed 15 bar oxygen pressure, as used in the nominally grown La$_{2.4}$Zr$_{0.6}$Ni$_{2}$O$_{7}$, and we gradually increased the pressure up to 150 bar, where the growth remained stable. We note that the application of 300 bar pressure led to growth instabilities. 

\begin{figure}[tb]
\includegraphics[width=1\columnwidth]{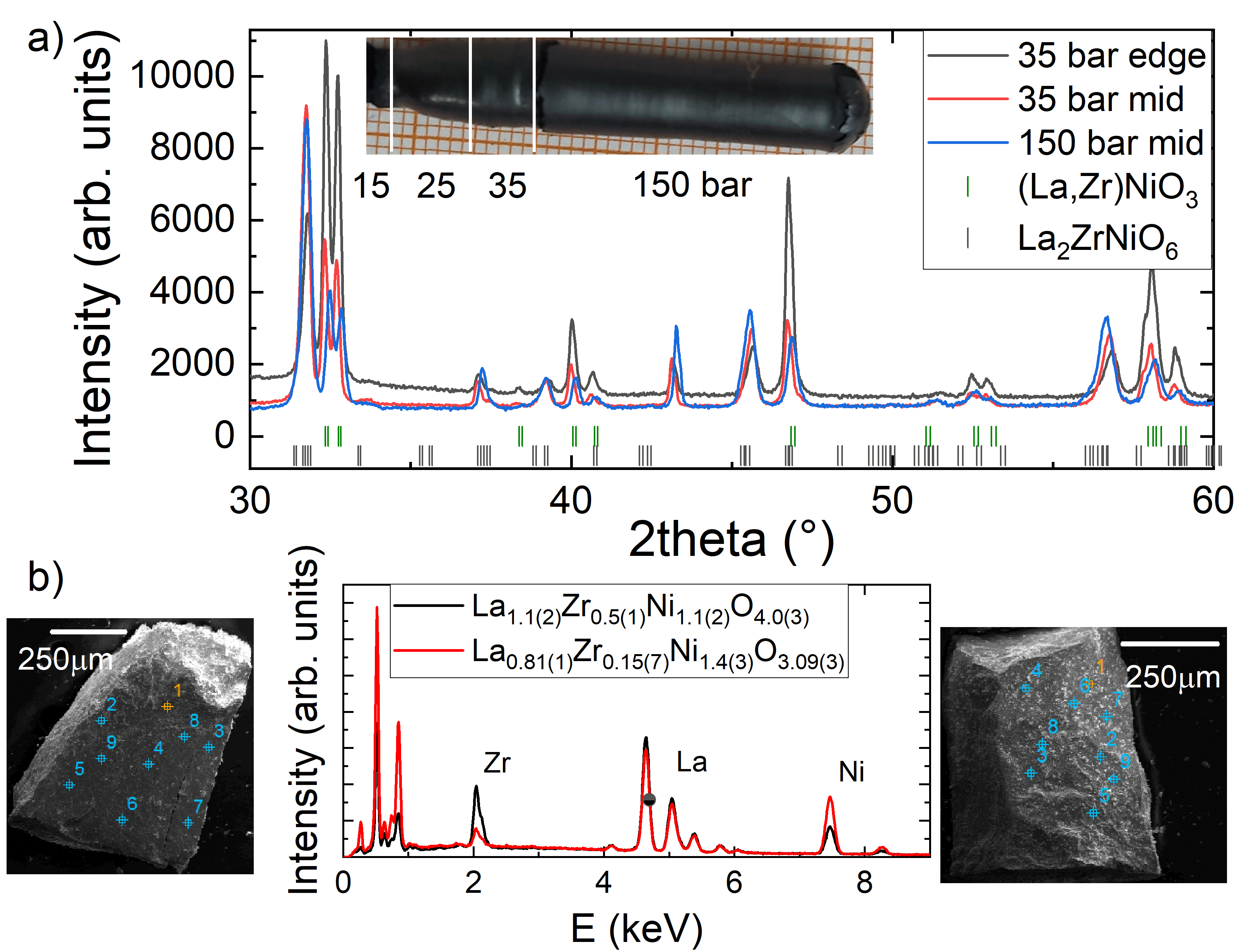}
\caption{(a) Image of the boule of a growth of the nominal composition La$_{0.8}$Zr$_{0.2}$NiO$_{3}$ using various oxygen pressures, together with the PXRD patterns from the edge region of the 35 bar part (black), the middle region (red), and the middle region of the 150 bar part (blue). The green bars indicate the peak positions of the perovskite phase from the refinement, while the gray bars indicate the double-perovskite phase. (b) The SE images feature broken pieces of the two separate phases, primarily consisting of $n=1$ Ruddlesden-Popper (left) and perovskite phase (right), and an exemplary  EDS result (center). The resulting average stoichiometry is given in the legend,  with La$_{1.1(2)}$Zr$_{0.5(1)}$Ni$_{1.1(2)}$O$_{4.0(3)}$ and  La$_{0.81(1)}$Zr$_{0.15(7)}$Ni$_{1.4(3)}$O$_{3.09(3)}$.}
\label{SM1}
\end{figure}

In Fig. \ref{SM1}, we present a summary of the pressure change effects, revealing that phase formation remains unaffected from 35 bar onwards. Consequently, we chose an intermediate pressure of 85 bar for our final growth. Although the BSE images display a significant phase mixture (Fig. \ref{LZNO}, we can still isolate small pieces of major phase accumulations from the $n=1$ Ruddlesden-Popper phases and perovskite phases. Interestingly, the perovskite parts appear to contain an increased substitutional content of Zr (Fig. \ref{LZNO}), which coexist alongside of tiny domains of NiO and the $n=1$-Ruddlesden-Popper phase.

\section{Elemental Dispersive X-ray Spectroscopy}

\begin{figure}[tb]
\includegraphics[width=1\columnwidth]{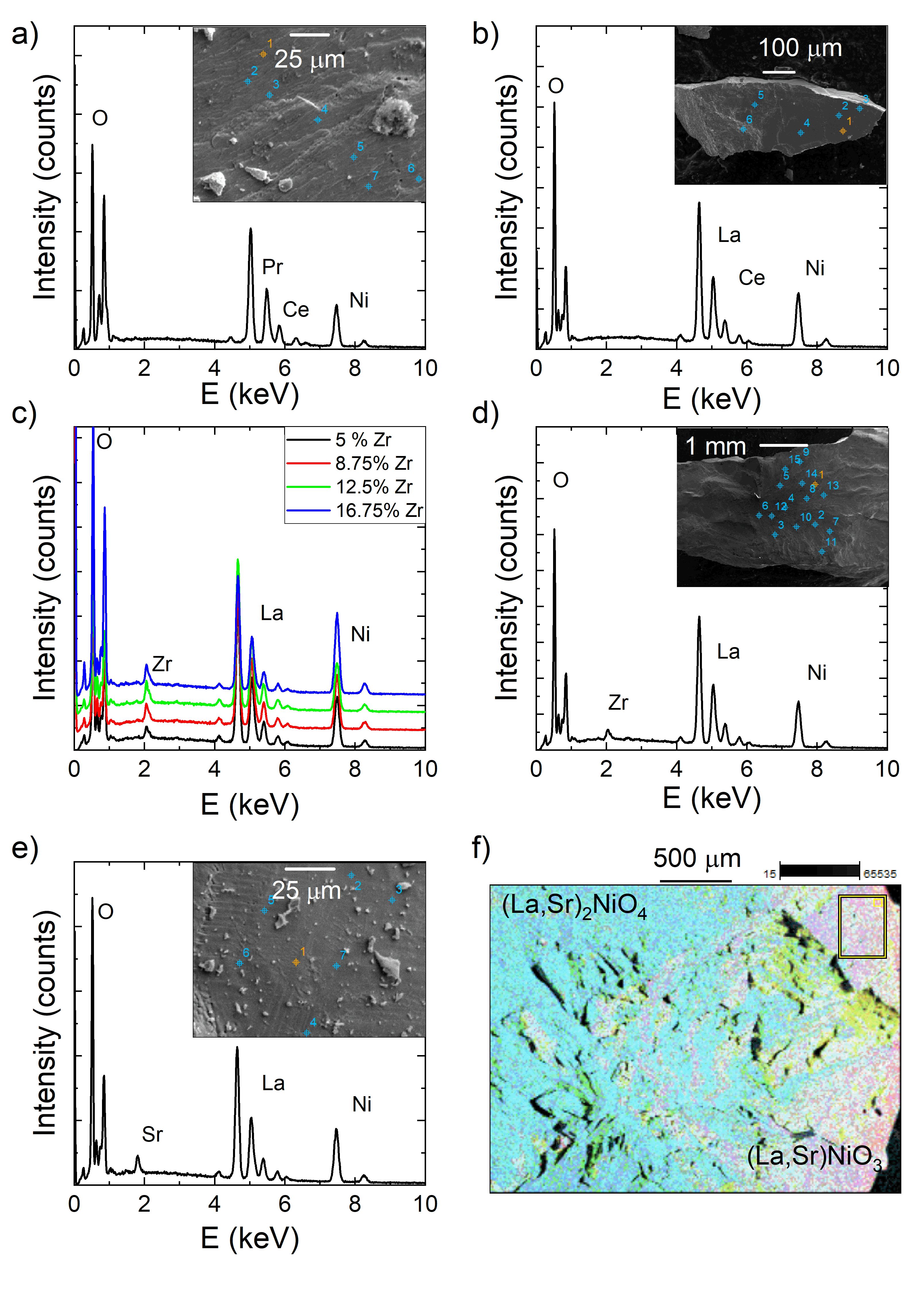}
\caption{EDS scans of crystals, with an exemplary SE image of a broken piece shown in the inset, for nominally grown (a) Pr$_{0.95}$Ce$_{0.05}$NiO$_{3}$, (b) La$_{0.91}$Ce$_{0.09}$NiO$_{3}$, (c) (La,Zr)NiO$_{3}$ (d) La$_{0.95}$Zr$_{0.05}$NiO$_{3}$,  (e) La$_{0.92}$Sr$_{0.08}$NiO$_{3}$. (f) EDS map of  La$_{0.84}$Sr$_{0.16}$NiO$_{3}$ of a "cleaved" cross section of the boule, with all elements overlaid in one picture on the SE picture, using the color code: La (red), Sr (purple), O (blue), Ni (green).}
\label{EDS}
\end{figure}

In addition to the extensive XRD and BSE analysis, we also performed EDS characterization, with representative results summarized in Fig. \ref{EDS}. We focused the EDS analysis on homogeneous pieces with lower doping levels, confirming the successful substitution through a  systematic examination of at least 20 points across various regions of the grown boule on the phase-pure part of perovskite matrix. An example SEM image is provided in Fig. \ref{EDS}.

By averaging over all measured points and considering the standard deviation as error bars, we obtained the following substitutional contents: (a) Pr$_{0.93(3)}$Ce$_{0.02(1)}$NiO$_{3}$, (b) La$_{0.96(2)}$Ce$_{0.046(3)}$NiO$_{3}$, (c) 5\%: La$_{0.99(9)}$Zr$_{0.071(4)}$NiO$_{3}$, 8.75\%: La$_{0.96(5)}$Zr$_{0.11(1)}$NiO$_{3}$, 12.5\%: La$_{0.9(2)}$Zr$_{0.12(2)}$NiO$_{3}$, 16.75\%: La$_{0.91(6)}$Zr$_{0.12(1)}$NiO$_{3}$, (d) La$_{0.9(2)}$Zr$_{0.07(2)}$NiO$_{3}$,  (e) La$_{0.8(1)}$Sr$_{0.12(5)}$NiO$_{3}$.

The phase separation, evident in both BSE and XRD data for doping levels of 8\% and higher, is due to the solubility limit of the substituent. This phenomenon is also reflected in the EDS data, which reveal a limited substitution content in the bulk matrix. It is important to note that, in comparison to the rare-earth elements, the Zr and Sr content is systematically overestimated in our EDS analysis by a few percent. The radial distribution of phases in the boule, resulting from the varying oxygen partial pressure within the melt, is best illustrated in our EDS map displayed in  Fig. \ref{EDS}f. In this figure, all elements are color coded and superimposed onto the same SE image, revealing a red perovskite region at the exterior and a blue $n=1$ Ruddlesden-Popper phase region at the interior of the grown boule. 

\section*{References}

\bibliography{aipsamp}

\end{document}